\documentclass[preprint,12pt]{aastex701}

\shorttitle{HOPS 373 Monitoring}
\shortauthors{Tobin et al.}

\newcommand{\kms}{\mbox{km s$^{-1}$}}
\newcommand{\microjy}{\mbox{$\mu$Jy}}

\newcommand{\lsun}{\mbox{L$_{\sun}$}}

\begin{document}

\title{Radio Continuum and Water Maser Monitoring of the Outburst in HOPS 373: Free-Free Emission Does Not Respond to the Outburst }
\author[0000-0002-6195-0152]{John J. Tobin}
\affiliation{National Radio Astronomy Observatory, 520 Edgemont Rd., Charlottesville,VA 22903, USA}
\email{jtobin@nrao.edu}
\author[0000-0002-6773-459X]{Doug Johnstone}
\affiliation{NRC Herzberg Astronomy and Astrophysics, 5071 West Saanich Rd., Victoria, BC V9E 2E7, Canada}
\affiliation{Department of Physics and Astronomy, University of Victoria, Victoria, BC V8P 5C2, Canada}
\email{}
\author[0000-0002-7154-6065]{Gregory Herczeg}
\affiliation{Kavli Institute for Astronomy and Astrophysics, Peking University, Yiheyuan Lu 5, Haidian Qu, 100871 Beijing, People’s Republic of China}
\affiliation{Department of Astronomy, Peking University, Yiheyuan 5, Haidian Qu, 100871 Beijing, People’s Republic of China}
\email{}

\author[0000-0003-3119-2087]{Jeong-Eun Lee}
\affil{Department of Physics and Astronomy, SNUARC, Seoul National University, 1 Gwanak-ro, Gwanak-gu, Seoul 08826, Republic of Korea}
\email{}
\author[0000-0002-3808-7143]{Ho-Gyu Lee}
\affiliation{Korea Astronomy and Space Science Institute, 776 Daedeok-daero, Yuseong, Daejeon 34055, Korea}
\email{}
\author[0000-0003-1894-1880]{Carlos Contreras Pe\~{n}a}
\affiliation{Department of Physics and Astronomy, Seoul National
University, 1 Gwanak-ro, Gwanak-gu, Seoul 08826, Korea}
\affiliation{Research Institute of Basic Sciences, Seoul National
University, Seoul 08826, Republic of Korea}
\email{}
\author[0000-0001-6216-0462]{Sung-Yong Yoon}
\affiliation{The School of Space Research, Kyung Hee University, 1732 Deogyeong-daero, Giheung-gu, Yongin-si, Gyeonggi-do, Republic of Korea}
\affiliation{Korea Astronomy and Space Science Institute, 776, Daedeok-daero, Yuseong-gu, Daejeon, 34055, Republic of Korea}
\email{}

\author[0000-0002-6956-0730]{Steve Mairs}
\affiliation{NRC Herzberg Astronomy and Astrophysics, 5071 West Saanich Rd., Victoria, BC V9E 2E7, Canada}
\email{}
\begin{abstract}
We present VLA C-band (5~cm) continuum, K-band (1.3~cm) continuum, and water maser (22.235 GHz) monitoring of the protostar HOPS-373. We additionally present the contemporaneous monitoring for 95 sources within the 5~cm field of view for over two years during the peak of the HOPS-373 outburst and an additional epoch in 2026. HOPS-373 is a binary Class 0 protostar located in the Orion star forming region that was found to have a $\sim$4$\times$ luminosity burst from the JCMT Transient Survey and NEOWISE monitoring. We do not find evidence for a change in the free-free emission traced by VLA 5~cm continuum during the peak of its outburst or during the decline. Moreover, the 1.3~cm continuum does not show significant variability between the NE and SW components of the HOPS-373 binary. The water maser emission is highly variable toward HOPS-373, multiple velocity components are detected at different (or the same) times and the maser spots are located close to the 1.3~cm continuum source of HOPS-373-SW. There is tentative evidence for the water maser spots to be propagating away from the source, but there is not a robust connection between the outburst and the observed maser activity.
The lack of correlation between outburst and free-free emission from HOPS-373 indicates that the free-free emission may not directly respond to increases in the accretion rate and subsequently the outflow rate. The lack of a link could be due to the outflow mostly being neutral, or there may be offsets in the timescale for the free-free response.
\end{abstract}

\section{Introduction}

The accretion of matter onto stars through a disk is expected to result in some expulsion of material through jets and outflows \citep{frank2014}. This phenomenon is predicted by theory and simulations of mass accretion and outflow driving for young stars (and more exotic objects like black holes and neutron stars)  \citep[e.g.,][]{blandford1982,pelletier1992,shu1994,matt2005}. As such, variations in accretion are also expected to result in variations in the outflow rate from protostars \citep{dunham2014, fischer2023}, which are expected to have observational signatures. 

In particular, the jets from protostars can often be punctuated with discrete knots of emission that can appear in infrared emission lines, CO rotation lines at submillimeter/millimeter wavelengths \citep{plunkett2015, watson2016, reipurth1989, ray2023, kim2024, dutta2024}, and free-free emission \citep{pech2010}. These knots are thought to be related to increases in the accretion rate that correspondingly increase the mass outflow rate, but  \citet{raga2015} alternatively suggest that there could be ejection time variability that is not specifically linked to accretion. A direct causal connection between an accretion variation (from an accretion luminosity increase) and an ejected outflow knot remains elusive due to the need to have sufficiently high resolution and observations before and after the burst to confirm that such outflow knots were indeed ejected during a burst. The knot in the jet of B335 \citep{kim2024} has the best evidence of being directly related to an outburst. An outburst was observed with the WISE/NEOWISE ([Near-Earth Object] Wide-Field Infrared Survey Explorer) survey data \citep{kim2024} and the estimated propagation velocity of the knot indicates that it was likely ejected around the time of the burst. 

While atomic and molecular lines are one way of tracing the outflow, they are observed at 10s to 1000s of au from the protostars and these features can take years to propagate to those locations. Thus, by the time a knot can be observed, the outburst event may have already concluded without being detected via photometry or otherwise noticed. Thus, an outflow tracer that can probe shorter timescales, closer to the driving source, is desirable for detecting a direct link between an outburst and an increased outflow rate. One such mechanism is free-free emission at centimeter wavelengths. Free-free emission is the direct result of internal shocks near the base of the protostellar jets and is detected frequently toward embedded protostars of all luminosities \citep{reynolds1986,anglada2018}. The free-free emission from protostars is strongly correlated with the mass outflow rate derived from molecular line observations as well as the bolometric luminosities of protostars over several orders of magnitude \citep{anglada2018}. Under the assumption that a substantial portion of protostellar luminosity comes from accretion across the range of luminosities from 0.1 to $\sim$10$^{5}$~\lsun\ \citep[c.f.,][]{hartmann2025}, the correlation between luminosity and free-free emission indicates a direct direct relationship between the accretion and outflow processes \citep{dunham2014,fischer2023}.

Free-free emission from protostars can be in the form of resolved jets \citep[e.g.,][]{anglada2018,tychoniec2018a}, but the vast majority of this emission is associated with compact point sources centered on or very near the dust continuum sources. Thus, free-free emission could be a useful tracer of the short timescale outflow rate from protostars. However, attempts to search for increased free-free emission from known repeating bursts, like the periodic pulsed-accretion binary protostar LRLL 54361 \citep{muzerolle2013}, did not show a signature of the burst in free-free emission \citep{forbrich2015}. And an attempt to detect a signature of the HOPS-383 burst \citep{safron2015} using archival data from years before the outburst to years after \citep{galvanmadrid2015} was inconclusive. Maser emission has also been utilized as a signpost of accretion bursts, but its utility has thus far been limited to high-mass protostars \citep[e.g.,][]{burns2016,brogan2018}, where masers directly respond to the varying radiation field.

Despite the inconclusive results thus far, a purposeful monitoring program to detect the signature of a burst and its decline in free-free emission would test its robustness as a short-timescale outflow tracer. Fortuitously, modest accretion bursts were detected toward the protostar HOPS-373 \citep{yoon2022} using data from the JCMT Transient Survey \citep{lee2021} and NEOWISE monitoring \citep{park2021}. There was an initial burst in late 2019 and it had faded by early 2020. Then in late-2020, when JCMT observations of the region had resumed, it had returned to the amplitude of the 2019 burst and then reached its highest 0.87~mm flux density around March 2021. Since late 2021, HOPS-373 has been steadily declining in 0.87~mm flux density toward its pre-burst state. At the peak of the outburst, the luminosity of HOPS-373 was inferred to have increased by $\sim$4$\times$ \citep{yoon2022}. During quiescent periods, the luminosity of HOPS-373 is 5.3~\lsun\ and has a bolometric temperature of 37~K \citep{tobin2020,furlan2016}.

We present follow-up observations of HOPS-373 at 5~cm (C-band) with the NSF's Karl G. Jansky Very Large Array (VLA) starting when HOPS-373 was near its peak in submillimeter flux density in 2021 March and we have monitored the source as it has faded in the submillimeter to search for a signature of the burst in centimeter-wave free-free emission. The VLA monitoring program with approximately monthly cadence was concluded in 2023 September, but a final epoch was obtained in 2026 to characterize any long-term response. We also monitored the water maser emission and 1.3~cm (K-band) continuum, toward the source, once in 2021 April, but with a 10 month gap, resuming in 2022 February. Furthermore, with the addition of pre-burst archival data from 2015 and 2016 in K- and C-bands, respectively, we have an ideal dataset to conclusively test whether there is a clear link between the free-free and/or water maser emission and the accretion burst that produced the submillimeter and NEOWISE increase in flux density. 

The paper is organized as follows. The observations are described in Section 2; the results are described in Section 3, we discuss the results in Section 4, and we present our summary and conclusions in Section 5.

\section{Observations and Data Reduction}
We used the VLA, located on the plains of San Agustin in New Mexico, USA to monitor the Class 0 protostar HOPS-373. We monitored the protostar at 5~cm continuum starting in 2021 March until 2023 September, with an additional observation in 2026 February. We also monitored the protostar at 1.3~cm continuum and the H$_2$O maser line at 22.235 GHz beginning in 2022 February and concluding in 2023 September, with additional observations in 2021 April and 2026 February. We also make use of archival data at 5~cm and 1.3~cm from 2016 October and 2015 November, respectively.

\subsection{C-band Observations and Data Reduction}
We observed HOPS-373 through two full cycles of all VLA configurations, starting with D-configuration in 2021 March, and we also observed one additional epoch in 2026 February. The programs 21A-409 and 21A-423 were submitted as director's discretionary time (DDT) proposals, while others went through the standard time allocation process.

In each observation, we used J0552+0313 as the complex gain calibrator and 3C147 as the absolute flux density and bandpass calibrator. The duration of each observation, including overheads was $\sim$1.3 hours, of which $\sim$45-50 minutes were spent on HOPS-373. The final observation in 2026, however, was 2$\times$ longer to reach sufficient S/N because only one observation was planned. We provide a condensed set of C-band observation details in Table \ref{vla_cband_obs}; we note that if execution blocks (EBs) were run within $\sim$1~day of each other, we imaged those together as a single epoch. We used the standard 3-bit continuum setup for observations through 2022 April, where the correlator is configured with 32 spectral windows, each with 128 MHz of bandwidth divided into 64 channels, sampling the parallel and cross-hand polarization products. This setup sampled the full 4 GHz of bandwidth available in C-band. Beginning with the A configuration observations in 2022 May, we began using a hybrid 3-bit/8-bit continuum setup because of strong radio frequency interference (RFI) at 6.2 GHz, likely due to microwave links that transmit across the array arms in the most extended configuration. These data showed signs of receiver gain compression, a reduction of the receiver response to input power as a result of strong RFI. The receiver gain compression is most severe with the 3-bit samplers, and our initial hybrid setup centered the 1 GHz 8-bit baseband at 7.4 GHz, providing additional bandwidth in cases where the entire 6 to 8 GHz portion of a 3-bit baseband needed to be flagged for certain antennas. This setup is denoted as 3-bit/8-bit-v1 in Table \ref{vla_cband_obs}.
Then for our final A-configuration observations in 2023 August-September and 2026 February, we made use a of a new spectral setup where the 8-bit baseband was tuned to 6.2 GHz and the 3-bit baseband was centered at 7.6 GHz. This setup uses the 1 GHz of bandwidth from the 8-bit baseband to cover the spectral range where the RFI may be present, since the 8-bit samplers can cope with more input power from interference without becoming non-linear, and the 3-bit baseband is tuned so that the RFI is out of the band, resulting in that baseband no longer being affected by gain compression. This setup is labeled as 3-bit/8-bit-v2 in Table \ref{vla_cband_obs}.

The data were reduced by the CASA VLA pipeline versions 6.1.2-pipeline-2020.1.0.36 (data prior to October 2021), 6.2.1-pipeline-2021.2.0.128 (data between 2021 October through 2022 September 20), 6.4.1-pipeline-2022.2.0.64 (data after 2022 September 20), and 6.6.6-pipeline-2025.1.0.363 (data taken in 2026). The VLA pipeline conducts bandpass calibration, flux density bootstrapping, complex gain calibration, and RFI flagging. Antennas affected by gain compression were flagged manually and the pipeline was re-run.

We also make use of two archival datasets from program 15A-369. These data were taken during a configuration change from A to D, but most antennas were in their D-configuration locations, enabling us to use this as a reference epoch of data for HOPS-373. The observations made use of the same setup as our standard 3-bit mode and the data were reprocessed using the CASA VLA pipeline version 6.1.2-2020.1.0.36, but we flagged the antennas in A-configuration prior to running the pipeline, in addition to scans and spectral windows affected by RFI.

The imaging was conducted manually using \textit{tclean} in CASA 6.1.2 or later. The imaging results were scientifically-equivalent between the different versions of CASA. The standard images we used in analysis for D-configuration images were generated with 1\arcsec\ pixels and were 2100$\times$2100 in dimension with a restoring beam of 14\farcs8$\times$11.4 and a 60.9\degr\ position angle, the C-configuration images used 0\farcs4 pixels and were 5400$\times$5400 in dimension with a 4\farcs9$\times$4\farcs9 restoring beam, the B-configuration images used 0\farcs125 pixels and were 8000$\times$8000 in dimension with a 2\farcs0$\times$1\farcs5 restoring beam and a 0\degr\ position angle, and the A-configuration images used 0\farcs075 pixels and were 12500$\times$12500 in dimension with a 0\farcs5$\times$0\farcs5 restoring beam. Each of these images used natural weighting and the multi-term multi-frequency synthesis (MTMFS) deconvolver with \textit{nterms} set to 2; thus, the imaging algorithm takes into account the spectral index of each component with the large fractional bandwidth in C-band. The images were cleaned using automatically generated masks using the \textit{tclean} \textit{auto-multithresh} masking \citep{kepley2020} with the \textit{sidelobethreshold} set to 2.0 and the \textit{noisethreshold} set to 3.75 and the threshold was set using the \textit{nsigma} parameter and set to 2.0. All images were also generated only using data from a $uv$-distance $>$ 1.4~k$\lambda$ in order to filter out large-scale emission from the nearby NGC 2068 nebula.

As shown in Table \ref{vla_cband_obs}, some of the data were self-calibrated. In general, the gains from self-calibration were minimal because decorrelation at C-band is not severe, even in A-configuration. The C-band analysis described in Section 2.3 also demonstrates that the self-calibrated A-configuration data do not have systematically larger flux densities than those in the first A-configuration where self-calibration was not applied. When self-calibration was applied, except for the manual self-calibration for the 2015 D-configuration data, we used automated self-calibration \citep{autoselfcal}.

\subsection{K-band Observations and Data Reduction}

The K-band monitoring of HOPS-373 was not as extensive as that of C-band due to going through the standard time allocation process instead of multiple DDT programs; however, 21A-409 was observed as a DDT program and was previously published in \citet{yoon2022}. The observations were conducted in much the same manner as those in C-band, with J0552+313 being used as the complex gain calibrator, 3C147 as the absolute flux calibrator, but we used 3C84 as the bandpass calibrator due to the need to perform bandpass calibration on spectral windows with higher spectral resolution. We provide a condensed set of K-band observation details in Table \ref{vla_kband_obs}.

The spectral setup for 21A-409 used the 3-bit samplers to provide nearly 8 GHz of bandwidth in 60 spectral windows that were 128 MHz width with 64 channels. Then, we used 4 spectral windows tuned to spectral lines. One was the water maser line at 22.235 GHz (16 MHz width, 512 channels), and the other three were tuned to the NH$_3$ (1,1), (2,2), and (3,3) inversion transitions, with 4 MHz bandwidth and 512 channels. The main focus of the observations in 21A-409 was the continuum, which is why the 3-bit samplers were used. 

The spectral setup for the subsequent observations used the 8-bit samplers with 16 spectral windows that are 128 MHz wide (64 channels) for measuring the continuum,  a 16 MHz spectral window with 512 channels tuned to the water maser line at 22.235 GHz, and 8 additional 4 MHz spectral windows tuned to CH$_3$OH transitions at 24.92871, 24.93347, 24.93438, 24.95908, 25.01812, 25.12487, 25.29442, 25.5414, and 25.802 GHz; however, the final spectral window was not tuned to the correct frequency for a CH$_3$OH line, but in any event, we do not discuss those CH$_3$OH observations further.

Finally, the spectral setup for the archival data (15B-229) also used the 8-bit samplers, but 14 spectral windows were 128 MHz wide with 64 channels and used for continuum, while five spectral windows were used for spectral line observations. The water maser at 22.235 GHz and CCS line at 22.34049 GHz were observed with 8 MHz windows having 512 and 1024 channels, respectively. Then the NH3 (1,1), (2,2), and (3,3) lines were also observed with 8 MHz windows and 1024 channels.

The data were reduced by the CASA VLA pipeline versions 6.1.2-pipeline-2020.1.0.36 (data prior to October 2021), 6.2.1-pipeline-2021.2.0.128 (data between 2021 October through 2022 September 20), 6.4.1-pipeline-2022.2.0.64 (data after 2022 September 20), and 6.6.6-pipeline-2025.1.0.32 (final data in 2026). Given that the default settings for the VLA pipeline does not always produce ideal results for spectral line data, we modified the pipeline in a few ways. First, for the 15B-229 data, we disabled the pipeline RFI flagging step and ensured that the CASA task \textit{statwt} only ran on line free portions of the line spectral window to avoid down-weighting spectral emission. Then for the 2021 April data and later, first backed up the flags prior to RFI flagging, allowed RFI flagging to run, and allowed \textit{statwt} to run as normal. This then properly sets the weights for the spectral line windows. We then restored the flags prior to RFI flagging using \textit{flagmanager}, split out the spectral line windows using \textit{mstransform}, and then reverted the flags back to their state following the execution of \textit{statwt}. This process resulted in two measurement sets with proper weights, one RFI flagged for continuum measurements and the other not RFI flagged for spectral line measurements. The K-band continuum is faint enough such that we did not need to perform continuum subtraction on the line data and the continuum data were too faint for self-calibration.

The resultant spectral cubes were generated with \textit{tclean} using CASA 6.1.2 or later.  The data from 15B-229 and 21A-409 were imaged with 512$\times$512, 0\farcs25 pixels and 170, 0.5~\kms\ channels, starting at -40~\kms; the earlier observations used more a narrow bandwidth around the water maser line than subseqent observations. The remaining data were all imaged with 200, 0.5~\kms\ channels, starting at -50~\kms; the BnA and A-configuration images used 512$\times$512, 0\farcs02 pixels, the B-configuration images used 1024$\times$1024, 0\farcs025 pixels, the C-configuration images used 1024$\times$1024 0\farcs125 pixels, and the D-configuration images used 512$\times$512, 0\farcs125 pixels. There were not other sources of emission within the primary beam outside the imaged region, enabling us to use the smaller angular sizes for images in some configurations. Continuum images in each configuration made use of the same pixel sizes and dimensions. Natural weighting was used for the spectral cubes and continuum images were made with different values of robust with Briggs weighing. Each image was cleaned down to the 3$\sigma$ level using the \textit{tclean} parameter \textit{nsigma=3.0}, then each image (and spectral cube) used a manually defined clean mask that encompassed the source and the expected location of water maser spots.

\subsection{C-band Continuum Analysis}

The data collected toward HOPS-373 at 5~cm totaled $\sim$54 hours (with overheads), amounting to $\sim$34 hours on-source toward the region. Due to the large primary beam in C-band ($\sim$11\arcmin, full width 10\% maximum), we will detect many more sources than just HOPS-373. Several HOPS sources were present within the 20\% power point of the primary beam: HOPS-373, HOPS-403, HOPS-321, HOPS-324, HOPS-325, HOPS-363, and there were additional HOPS sources between the 10\% and 20\% power point: HOPS-320, HOPS-388, HOPS-389, and HOPS-322. In addition to the protostars classified by \citet{furlan2016}, there were also many more-evolved YSOs and extragalactic sources contained within the field of view; some are found in other catalogs but many others are not. Thus, we take two approaches for measuring the flux densities of the sources in the images. First, since we know the location of the detected HOPS sources, we used manually drawn regions, to restrict the fitting area, and used the CASA task \textit{imfit} to fit Gaussians to the known sources and measure their flux densities, using the primary beam corrected images. HOPS-373 is a known binary with $\sim$ 4\arcsec\ separation, so we fitted it with a single Gaussian for the D-configuration images and two Gaussians for all the other images and sum the flux densities of the two components. This approach enables us to better measure the flux densities for faint sources that would not always be detected by automated source finders.

Then to measure the flux densities of all the detected sources in an automated, unbiased manner, we made use of pyBDSF \citep{pybdsf2015}, and ran it on the primary beam corrected images. With the output of pyBDSF for each image, we characterized the accuracy of our flux density calibration across configurations and time. For A, B, and C configuration data, we also make images with the standard restoring beam for images in directly preceding lower-resolution configuration. For example, when the observations were conducted in A-configuration, we also made an image with the resolution of B-configuration using tapering and specifying the restoring beam, and did the same for other configurations. As such, we also combined all the visibility data taken during an array configuration to make a higher sensitivity, time-averaged image spanning several months. We then used the sources found in each of these time-averaged images to characterize differences in the observed source flux density distribution between configurations. 

To characterize the differences in the distribution of flux densities, we cross-matched the catalogs for each whole configuration image at the same resolution, we then took the ratio of the flux densities of matching sources and fit a Gaussian to the histogram of their flux ratio distributions. The expectation is that if sources are on average the same flux density and only randomly varying, as such the distribution should be centered at 1.0. We were able to do this for the first visits to D, C, B, and A configuration since we could make sufficiently sensitive images at the resolution of the preceding configuration. We show the results of this analysis in Table \ref{relative-scales}, where it can be seen that the sources fluxes from D to B configuration differ by $<$10\%, but the A-configuration source fluxes are systematically lower than the D-configuration by $\sim$17\%; see Table \ref{relative-scales}, column 4.

Then, because we cannot create a reasonably sensitive image at D-configuration resolution using A-configuration data, we cannot continue the comparison of subsequent consecutive configurations in time, but we can compare to the previous data in the same configuration. We see that during the second visit to D-configuration, its sources are slightly fainter with respect to the first D-configuration, while C and B-configurations are slightly brighter and then A-configuration is slightly fainter again. The final A-configuration observation in 2026 is slightly brighter than the 2022 and 2023 observations. This analysis demonstrates that our overall flux density scale is not systematically different by more than 10\% in D, C, and B configuration, but there could be systematically lower flux densities measured in A-configuration. Those systematics in A-configuration could be due to spatial filtering. This can happen if the more constant sources are marginally-resolved, i.e., free-free emitting YSOs, then their flux densities might not be recovered as well in A-configuration. However, we did attempt to mitigate the issues associated with spatial filtering by applying a uvtaper and using the same restoring beam as the lower-resolution configuration image. Despite some of the potential systematic differences between configurations shown in Table \ref{relative-scales}, we did not attempt to to apply any corrections given that the apparent differences could also result from small number statistics and there are many sources that appear approximately constant in flux density vs. time, e.g., HOPS-403 (Section 3.2).

We then use the average, whole configuration images to look for systematic variations during a particular configuration by cross-matching the pyBDSF source catalogs and examining the distribution of flux density ratios for the cross-matched sources. The results are shown in Table \ref{vla_cband_obs}, column 9, and it is clear that most observations within a configuration have flux density distributions that are consistent within 10\%, with a few outliers having higher variations.

\subsection{K-band Continuum Analysis}

The 1.3~cm data have a much smaller primary beam ($\sim$2\arcmin) and therefore only detect HOPS-373 in continuum and maser emission, making their analysis more straight forward. However, it is not possible to examine the data for possible systematic offsets in the flux density scale using a large number of sources. The 1.3~cm continuum provides ancillary information on the response of the circumstellar dust to the outburst because its emission is expected to be a combination of dust and free-free continuum. While its is known \textit{a priori} that the southwest source is the one that varied \citep{yoon2022}, we are able to continue to examine the absolute and relative flux densities of the sources at 1.3~cm and conduct a more rigorous analysis of any potential variability at 1.3~cm. We measured the flux densities of each source using Gaussian fits to the images from each epoch of observation, in addition an overall image generated using the emission from all observations in a given configuration.

\subsection{Water Maser Analysis}

To search for water maser emission we examined the data cube channel-by-channel, looking for point-like emission above the noise and near the continuum source(s). For each maser spot at a particular velocity and image, we recorded the approximate position and velocity to use as priors for later analysis. We then then extracted a spectrum toward the maser spots from each cube, sampling the total flux within a circular area, having a radius equivalent to the major axis of the synthesized beam. We then fit Gaussians to each detected water maser spectral component to derive the integrated flux density, velocity and line width for each component. Next, we extracted integrated intensity and peak intensity maps for the ranges of channels associated with each spectral component of water maser emission. Then to determine the position of each maser component, we fit Gaussians in the image plane to the peak intensity map.

\section{Results}

\begin{figure}
\begin{center}
\includegraphics[scale=0.6]{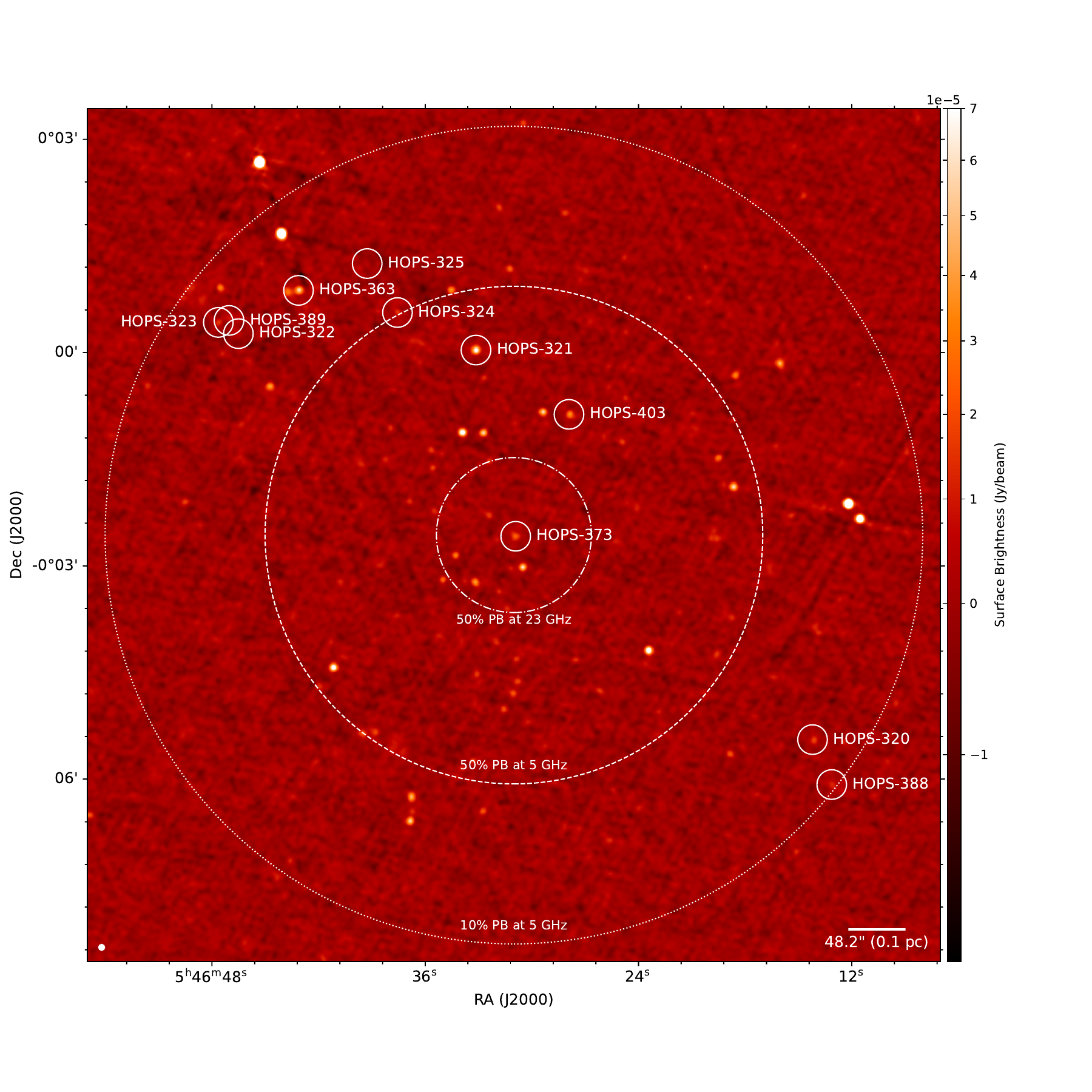}
\end{center}
\caption{Overview of the region around HOPS-373 as viewed at 5~cm by the VLA in C-configuration. This image is a combination of all C-configuration data, providing a good compromise between resolution and feature size for an overview of the region. The 5~cm 50\% and 10\% power points of the primary beam are marked with dashed and dotted circles, respectively. The inner dot-dashed circle delineates the 50\% power point of the 1.3~cm primary beam. All HOPS sources within the 10\% power point of the primary beam are labeled. The RMS noise in this map is 1.2~\microjy, and the beam size is 4\farcs9. This is a `flat noise' image where the response of the primary beam has not been corrected.
}
\label{HOPS-373-overview}
\end{figure}

 We show an overview image of the region at 5~cm in Figure \ref{HOPS-373-overview} at a resolution of 4\farcs9 using C-configuration with a sensitivity of $\sim$1.2~\microjy. The 1.3~cm data have a significantly smaller field of view and only detect HOPS-373, as described in the following sections. Thus, with the large field of view at 5~cm, we have the opportunity to study more sources that fall within our field of view. We will primarily discuss the results toward HOPS-373 and a few other well-detected HOPS sources that were also detected in the JCMT Transient Survey or have particularly interesting morphology in the VLA data. The other YSOs, including those that are more-evolved, have their results presented in Appendix A. Most more-evolved YSOs are likely tracing highly variable gyrosynchrotron emission \citep[e.g.,][]{dzib2013}. Variation in gyrosychrotron emission is not related to changes in luminosity and is of less direct relevance to the main topic of correlating any changes in dust and/or free-free and dust emission probed at 5~cm and 1.3~cm.

\subsection{HOPS-373 5~cm Continuum}

\begin{figure}
\begin{center}
\includegraphics[scale=0.6]{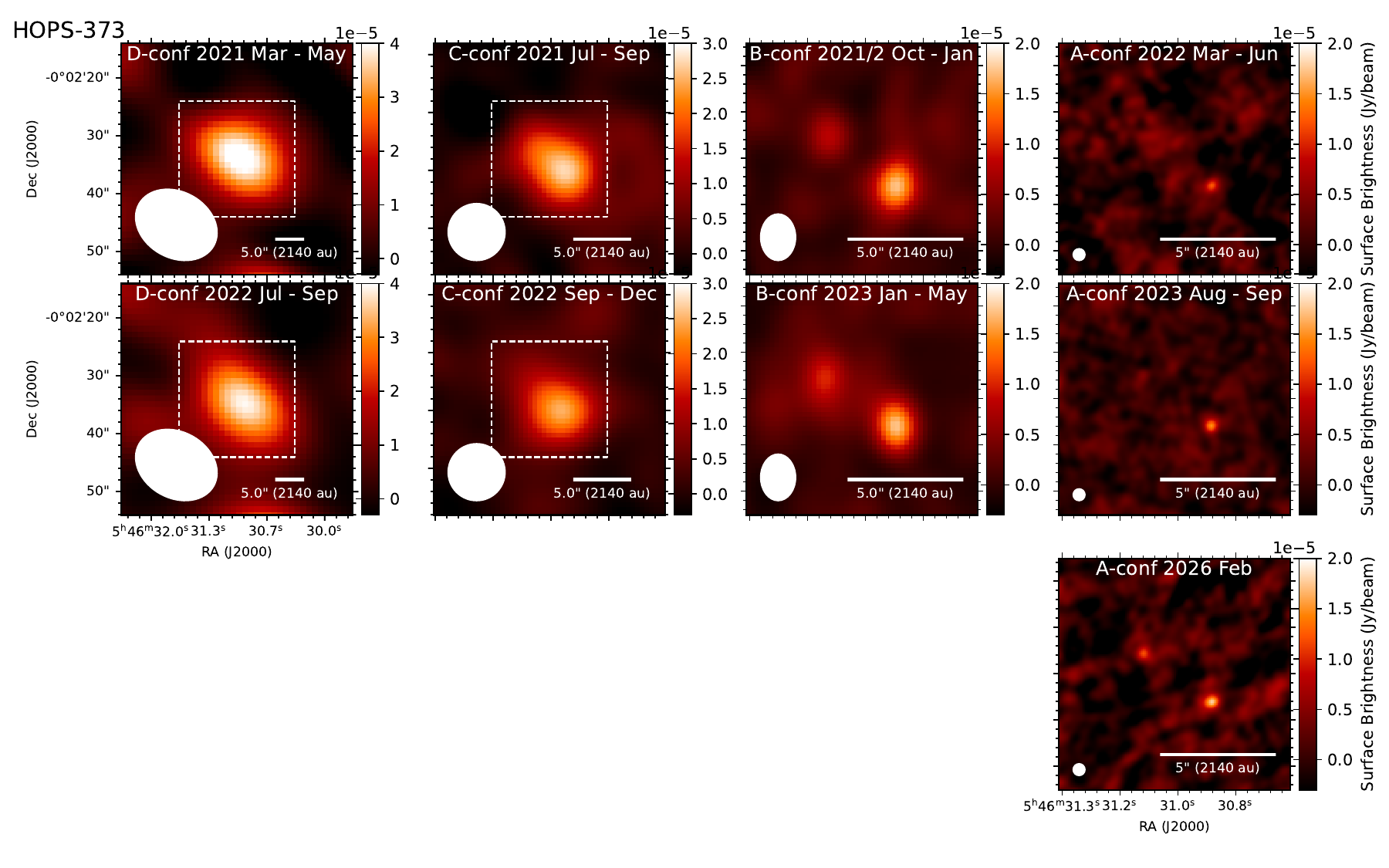}
\end{center}
\caption{Images of HOPS-373 from the VLA at 5~cm generated using all the data taken in a particular configuration. We show D-configuration (left), C-configuration (left-middle), B-configuration (right-middle), and A-configuration (right). The images from the first pass through the VLA configurations are shown in the top row, the second pass through are shown in the middle row, and the final observation is shown in the bottom panel. The time ranges for the observations included are annotated in each panel. The dotted boxes shown in the D- and C-configuration images denote the area zoomed-in on in the images shown from the higher resolution configurations. The beams are drawn in the lower left corner and a 5\farcs0 scalebar is drawn in each panel.
}
\label{HOPS-373-C-band}
\end{figure}

We show the average, per-configuration images, on the same color stretch for HOPS-373 in Figure \ref{HOPS-373-C-band}. The images show the clear change in morphology with angular resolution from a single blended source in D-configuration to a well-resolved binary source whose morphology appears point-like in A-configuration. We show the flux density over time for HOPS-373 as a whole, and HOPS-373~SW, and HOPS-373~NE in Figure \ref{C-band-vs-time-H373}, and with the JCMT lightcurve plotted below in Figure \ref{C-band-vs-time-wJCMT}. HOPS-373~SW and NE were not resolved from each other in D and C-configurations, thus, only their combined, total flux density is relevant in those configurations and SW and NE have upper limits plotted in Figure \ref{C-band-vs-time-H373}. The fact that NE is fainter than SW is also apparent in the figure given that NE was not detected in all B and A-configuration observations.

The lightcurves clearly shows that HOPS-373 did not show an increase in 5~cm flux density following the luminosity burst as traced by NEOWISE and the JCMT, nor did the flux density appreciably change from the pre-burst observations in 2015 through the end of monitoring in 2023. If we only consider the C and B-configuration observations, we can say that the flux density does not vary by more than 40\% throughout our monitoring. Nor is there evidence of a systematic change in the overall flux density as a function of time. Thus, the 5~cm continuum does not appear to show a response to the burst seen at submillimeter and infrared wavelengths.

The systematic drops in flux density in the higher-resolution configurations (A and B) are likely the result of spatial filtering where some weak, extended continuum emission around each (and both) protostars that is not fully recovered. Thus, the 5~cm emission around HOPS-373 is modestly extended and not simply two point sources. Further evidence can be seen visually in Figures \ref{HOPS-373-C-band} and \ref{C-band-vs-time-H373} where HOPS-373 NE is not well-detected in the A-configuration images relative to HOPS-373 SW. The peak intensities and integrated flux densities of SW are $\sim$2$\times$ greater than that of NE. The individual flux density measurements for HOPS-373 are compiled in Table \ref{per-epoch-imfit-fluxes}.

We do note, however, that the scatter in flux density is considerably larger in the B and A configurations at MJD $>$60000 ($\sigma_B$ = 7.82~\microjy, $\sigma_A$ = 8.64~\microjy ) vs. the prior observations in A and B configurations ($\sigma_B$ = 5.75~\microjy, $\sigma_A$ = 3.87~\microjy), despite a similar on-source time for all observations. HOPS-403 (discussed further in the next section) did show some increase in the scatter of the B and A configuration points at MJD $>$60000, but not as much as HOPS-373. However, given the faintness of the individual components of HOPS-373, it seems more likely that the increased scatter simply comes from measurement uncertainty and not a change in the properties of the source.

\subsection{HOPS 403 5~cm Continuum}
\begin{figure}
\begin{center}

\includegraphics[scale=0.75]{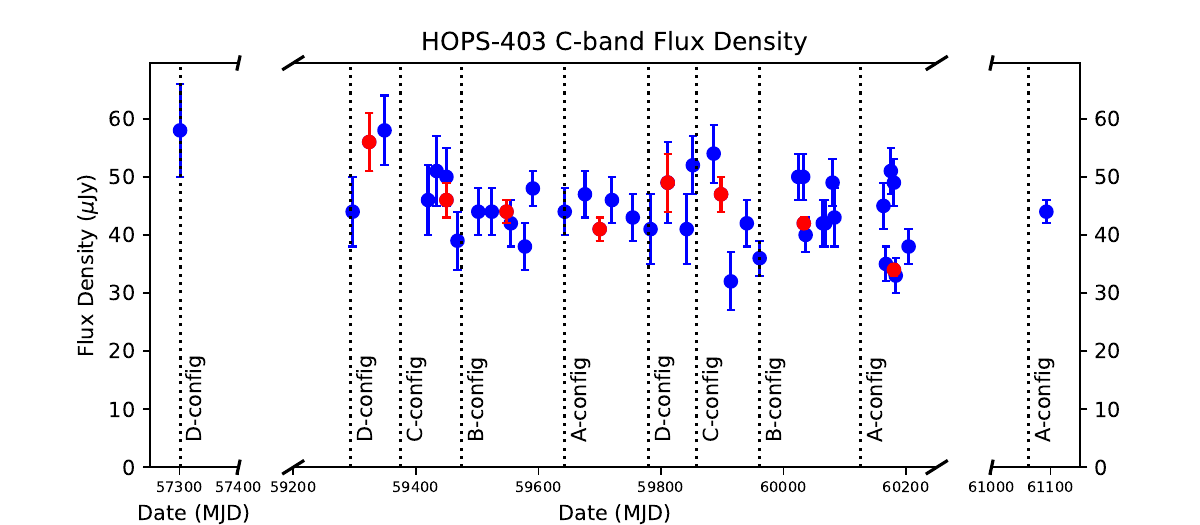}
\includegraphics[scale=0.75]{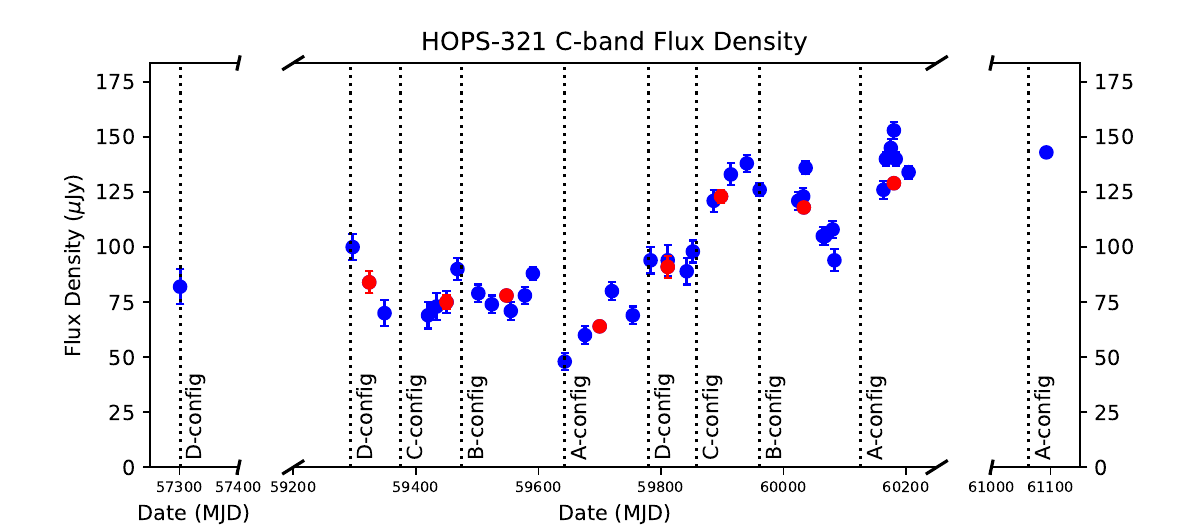}
\end{center}
\caption{The flux densities of HOPS-403 are quite constant with time, with little systematic change in different configurations. The HOPS-321 data show that its flux density increased by $\sim$1.5$\times$ in September 2022 and sustained this greater flux density in the higher resolution configurations until the end of monitoring in 2026 February. This is consistent with the apparent change in flux seen in the images shown in Figure \ref{HOPS-321-C-band}.
}
\label{C-band-vs-time-others}
\end{figure}

\begin{figure}
\begin{center}
\includegraphics[scale=0.6]{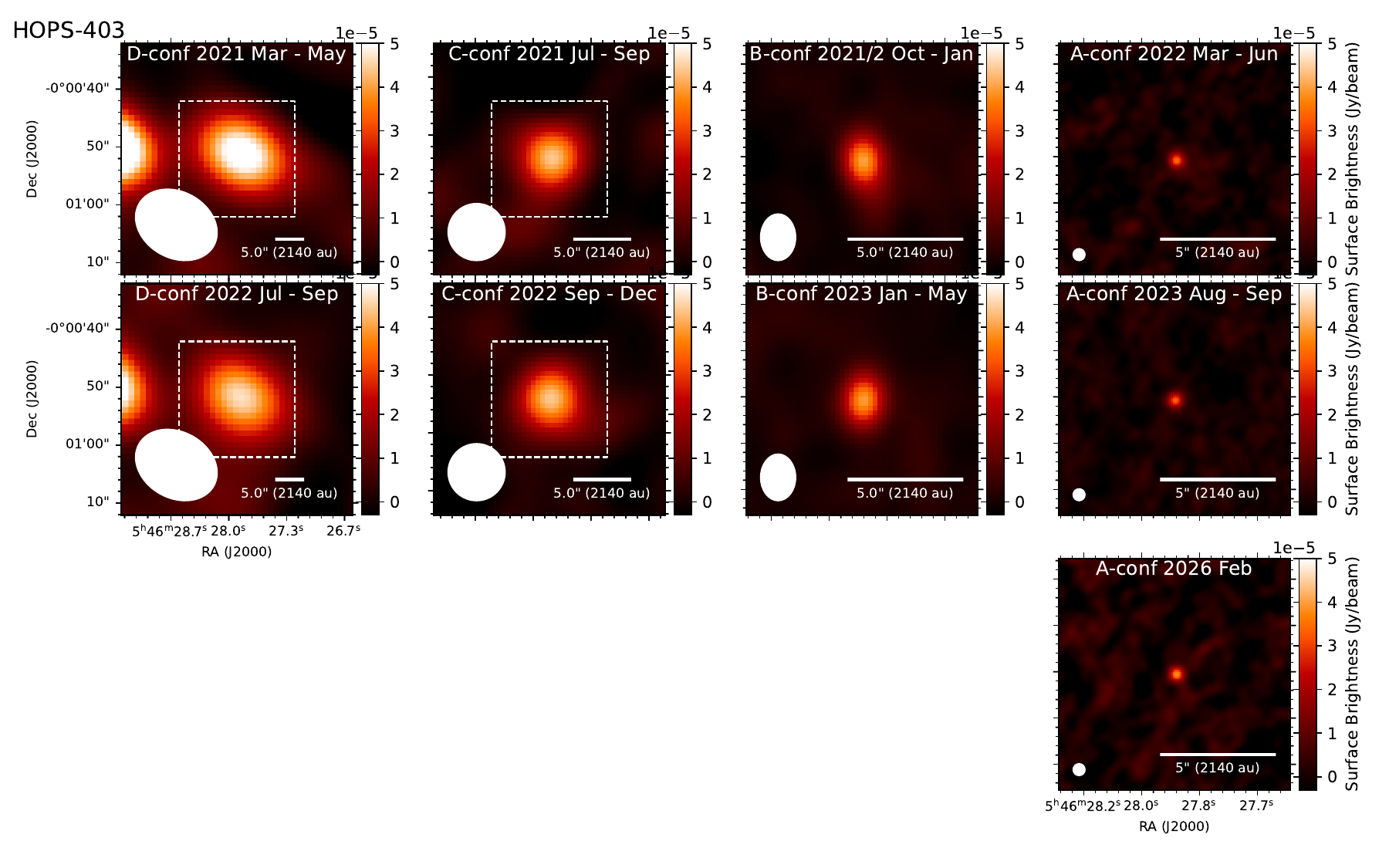}
\end{center}
\caption{Same as Figure \ref{HOPS-373-C-band} but for HOPS-403.
 The D-configuration image also shows an unrelated source at the east edge of the frame, likely a background galaxy.
}
\label{HOPS-403-C-band}
\end{figure}

HOPS 403 is located $\sim$1.8\arcmin\ NW of HOPS 373 at the 83\% power point of the primary beam. HOPS 403 is also an extremely young Class 0 protostar \citep{stutz2013,tobin2015}, like HOPS 373, and has a luminosity of 4.1~\lsun. We show the average, per-configuration images, on the same color stretch for HOPS 403 in Figure \ref{HOPS-403-C-band}. In contrast to HOPS 373, HOPS 403 remains point-like at all configurations and when imaged. The plot of flux versus time (Figure \ref{C-band-vs-time-others}) shows that the 5~cm flux density of HOPS-403 has remained nearly constant from 2015 to 2023, across all configurations. The JCMT flux, while not significantly varying, does seem to rise slightly toward the end of the overlapping JCMT monitoring period (Figure \ref{C-band-vs-time-wJCMT}), but no change in the 5~cm flux density is observed in the final observation. The flux density measurements for HOPS-403 are compiled in Table \ref{per-epoch-imfit-fluxes}.

\subsection{HOPS 321 5~cm Continuum}

\begin{figure}
\begin{center}
\includegraphics[scale=0.6]{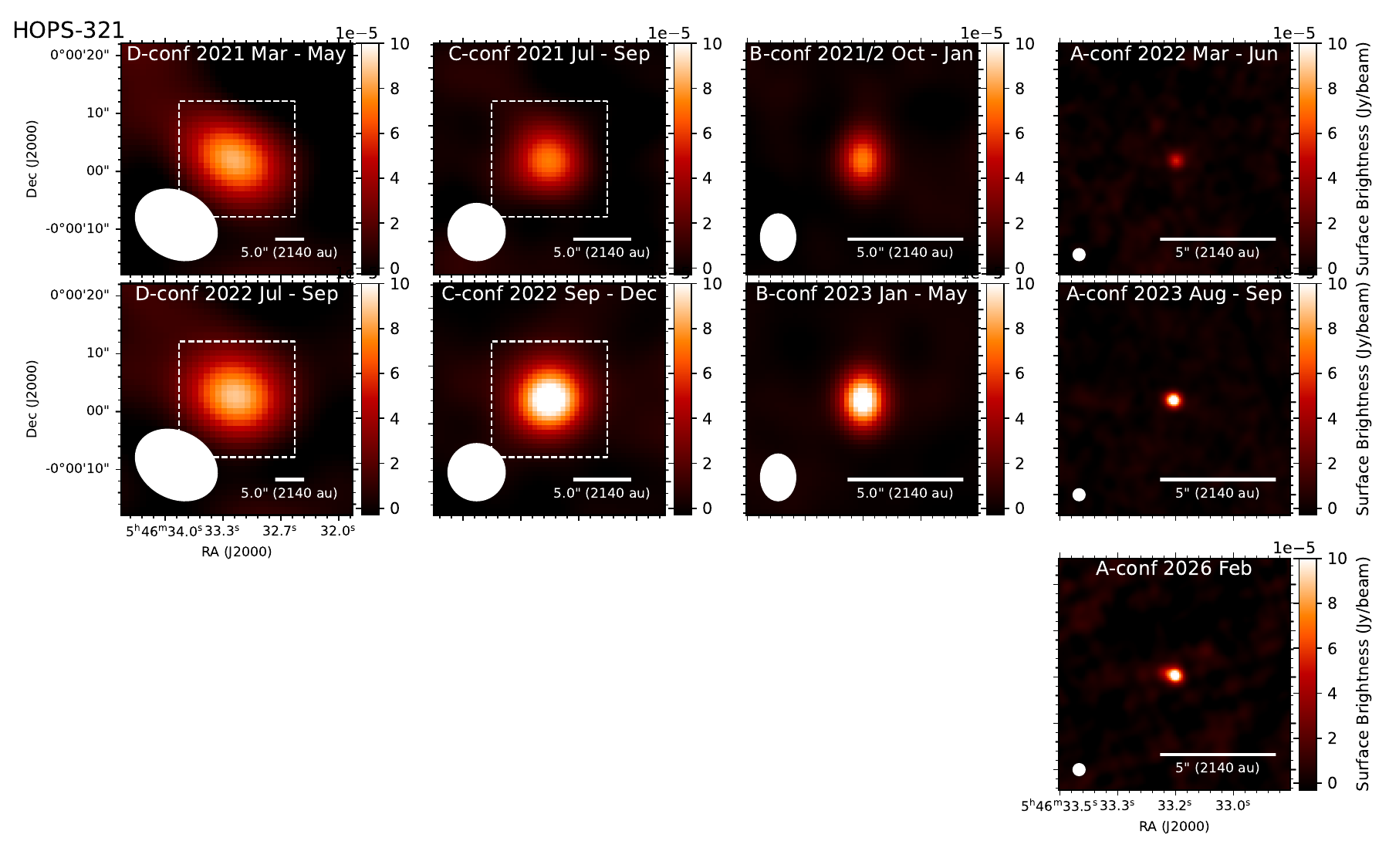}
\end{center}
\caption{Same as Figure \ref{HOPS-373-C-band} but for HOPS-321. The images starting in 2022 September show that HOPS-321 appears brighter due to an increase in its flux density.
}
\label{HOPS-321-C-band}
\end{figure}

HOPS 321 is located $\sim$2.5\arcmin\ NE of HOPS 373 at the 68\% power point of the primary beam. It is a Class I protostar with a luminosity of 3.7~\lsun\ and was marginally detected in the dust continuum with ALMA by \citet{tobin2020}. We show the average, per-configuration images, on the same color stretch for HOPS 321 in Figure \ref{HOPS-321-C-band}. The source shows clear differences in brightness at later times, and the final 5~cm image from 2026 February appears slightly extended in the eastern direction. The plot of flux versus time (Figure \ref{C-band-vs-time-others}) shows that the 5~cm flux density of HOPS-321 has some variations within a configuration, but the most robust feature is the overall increase in flux from $\sim$70-90~\microjy\ to $\sim$125-150~\microjy, shortly after the change to C-configuration at the end of 2022 September. This elevated flux density was still present in 2026 February. However, no corresponding signifcant change was seen in monitoring data from the JCMT (Figure \ref{C-band-vs-time-wJCMT}). The flux density measurements for HOPS-321 are compiled in Table \ref{per-epoch-imfit-fluxes}.

We examined the spectral index of the emission in the combined data from the two B-configuration observations, one in late 2021 through early 2022 and the other in 2023. These two periods sample the emission from the source before and after the flux density increase, respectively. The combined images needed to be used because HOPS-321 is near the edge of the primary beam and we need to make images of the upper and lower 2~GHz portions of the 5~cm bandwidth and measure the flux densities in those two images and calculate their spectral indices. The spectral index in the 2021/2022 data (pre-increase) was 0.82$\pm$0.13 and the spectral index in 2023 (post-increase) was 0.66$\pm$0.05. Thus, the spectral indices of the emission before and after the flux density increase are consistent within their uncertainties, and the positive spectral spectral index with increasing frequency indicates that the primary emission process is free-free.

\subsection{HOPS 363 C-band Continuum}

HOPS-363 is located $\sim$5\arcmin\ NE of HOPS-373 at the 29\% power point of the primary beam. It is a Flat Spectrum protostar, with a bolometric luminosity of 22.5~\lsun\ \citep{furlan2016,tobin2020}. The source in D-configuration clearly appears non-point like and is resolved into two components in C-configuration. The second component is still well-defined in the B-configuration images, but is less prominent in the A-configuration images as A is more highly-resolved. ALMA data toward HOPS-363 from \citet{tobin2022} resolve HOPS-363 into a compact binary system, but the companion is separated by 0\farcs185, unresolved in the 5~cm continuum presented here. The brighter source in the center of the images is presumably the protostar from its association with the ALMA dust continuum and infrared images. The ALMA CO ($J=3\rightarrow2$) data do reveal that the outflow of HOPS-363 is aligned with the orientation of the eastern resolved source, even though there is no strong molecular emission associated with it. Furthermore, the lower peak intensity in the A-configuration images suggest that this source is resolved and could be a shock in the outflow of HOPS-363. We note, however, that we did not observe any significant motion of this source during our monitoring, but 100~\kms\ over a 1.25 year period would only result in a shift of 0\farcs06, below our limit to detect with any confidence because the source is quite diffuse in A-configuration.

We also examined the spectral index of the emission, using the same strategy as for HOPS-321 in the preceding Section. However, HOPS-363 is even closer to the edge of the primary beam than HOPS-321, and will result in larger uncertainties on measured spectral indices. Thus, we simply compared the relative intensities of the peak emission toward each source in a primary beam corrected images generated from the high and low-frequency ends of C-band. Both in 2021/2022 and 2023, a positive spectral index is observed for HOPS-363 and the source to its east, indicating that both are tracing free-free emission.

\subsection{HOPS-373 1.3~cm Continuum}

\begin{figure}
\begin{center}
\includegraphics[scale=0.55]{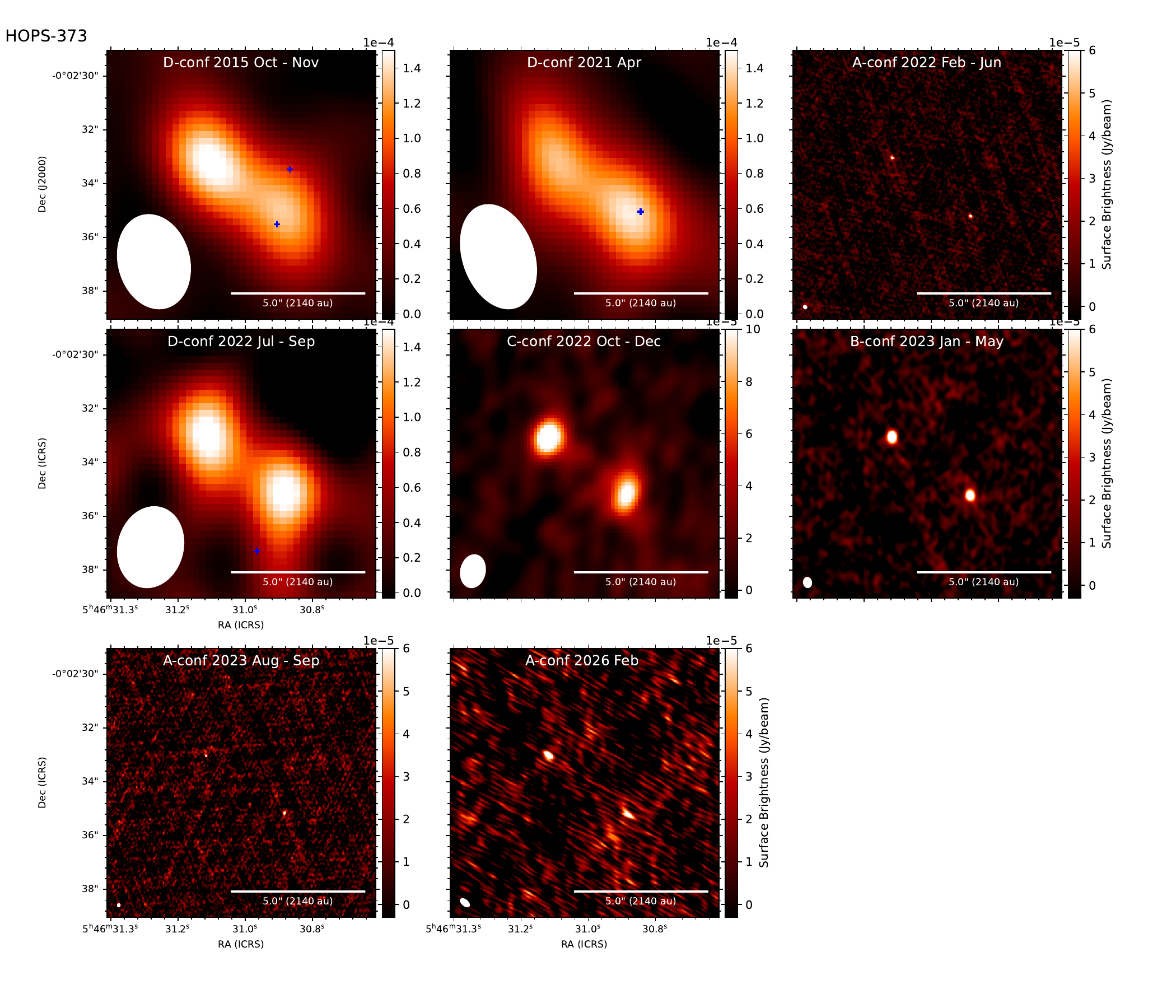}
\end{center}
\caption{Continuum images at 1.3~cm toward HOPS-373 using all the data taken during a particular configuration. The maser positions are overlaid on the D-configuration images for the masers detected in those configurations (no masers detected in C-configuration). The D-configuration images show, by-eye, that in 2021 the southwest source was brighter than the northwest source, reversing the pattern observed in 2015. Then in 2022 the sources appear equally bright in D-configuration again. The comparison of source flux densities and their ratios are consistent with no variability within their uncertainties (Figure \ref{HOPS-373-K-band-vs-time}).
}
\label{HOPS-373-K-band}
\end{figure}

\begin{figure}
\begin{center}
\includegraphics[scale=0.75]{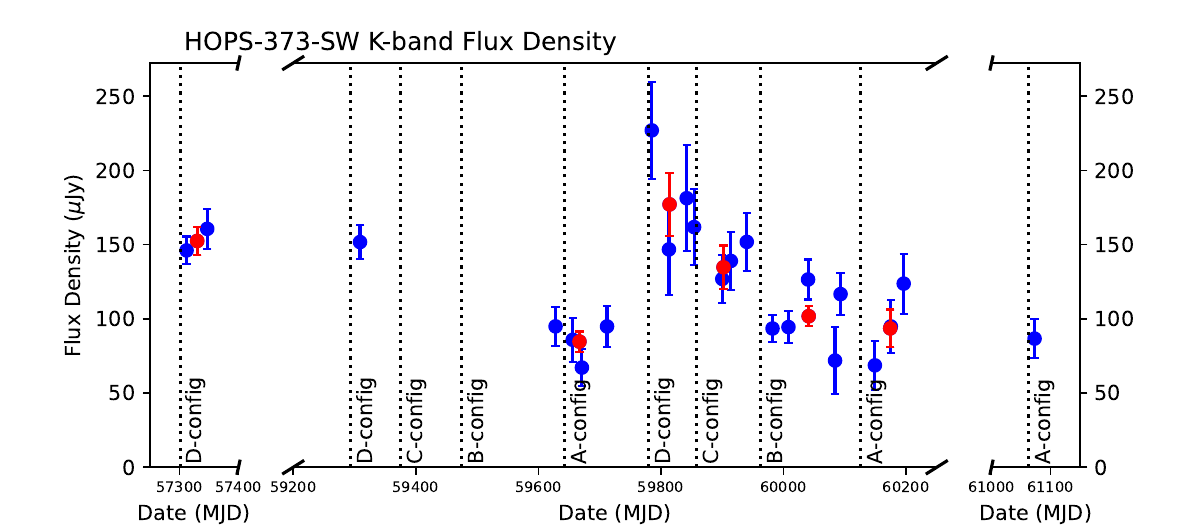}
\includegraphics[scale=0.75]{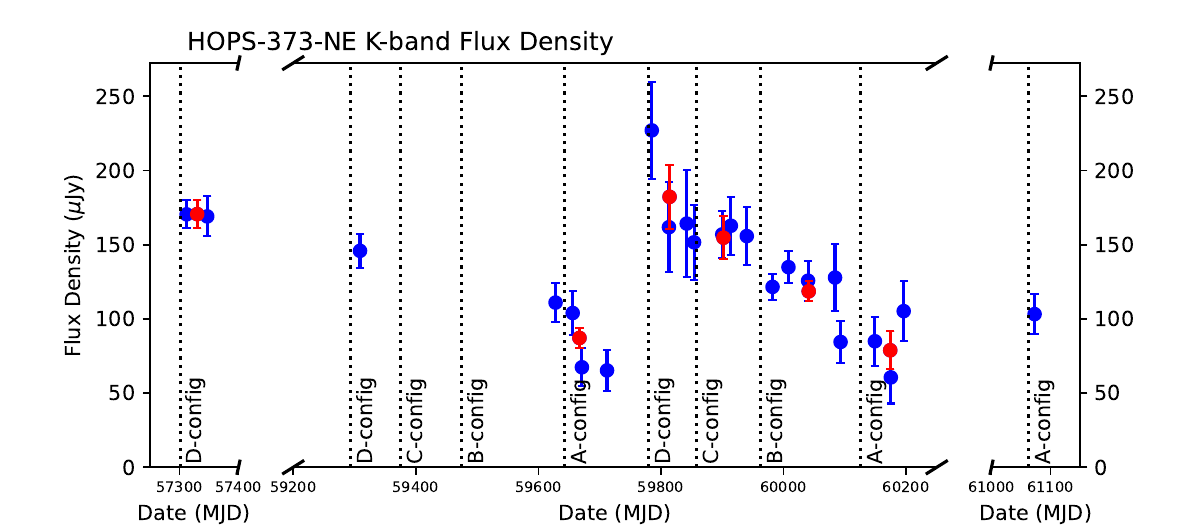}
\includegraphics[scale=0.75]{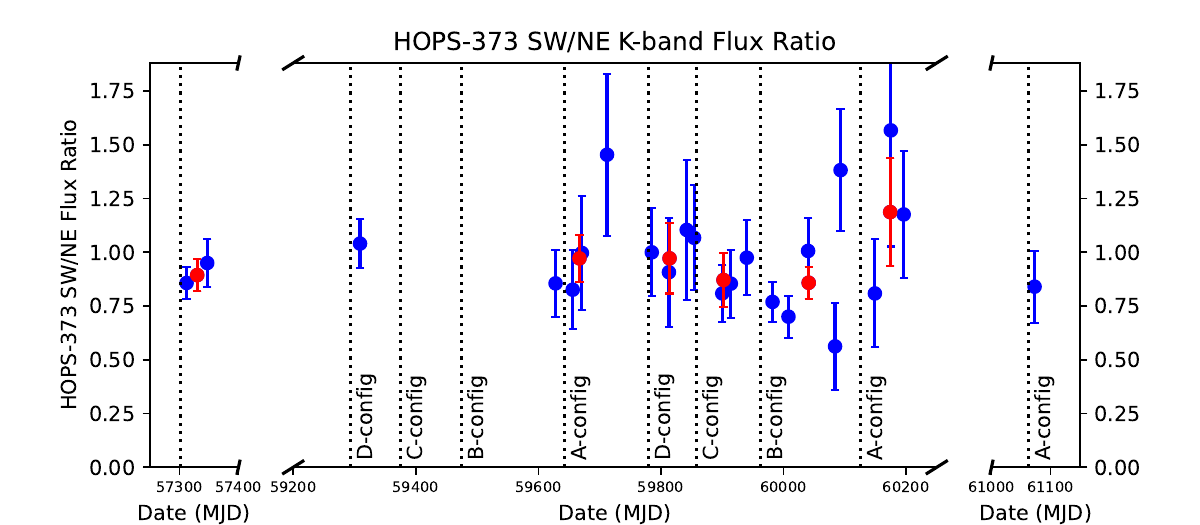}
\end{center}
\caption{Continuum flux density measurements at 1.3~cm toward HOPS-373 SW and NE (top and middle) and the ratio of their flux densities (bottom). Note that the observed flux density strongly depends on the configuration, resulting in the decline of flux density from D to A-configuration due to spatial filtering. There is no evidence for variability in the 1.3~cm continuum flux densities from either their absolute measurements or the ratio between sources.
}
\label{HOPS-373-K-band-vs-time}
\end{figure}

The 1.3~cm continuum is well-detected toward HOPS-373, and the two components of the system are resolved from each other in all configurations. We show the images from the combination of all the data taken in each visit to a particular VLA configuration in Figure \ref{HOPS-373-K-band}. We measured the 1.3~cm continuum emission toward each source for each observation and the flux densities vs. time for the sources are shown in Figure \ref{HOPS-373-K-band-vs-time}. The flux densities from D-configuration to A-configuration (2022 July to 2023 September) appear to decline, but this is due to spatial filtering, resulting in less overall flux being recovered with progressively higher resolution. As such, this is not evidence for a flux density decline associated with the fading burst. We can most directly compare the D-configuration observations from 2015 to the D- and C-configuration observations in 2021 and 2022 where the flux densities of both sources are consistent within their uncertainties. There is no statistically significant change in the flux densities within all the D- and C-configuration observations. We note that the flux density observed in the D-configuration observation on 2022 July 25 is $\sim$2.5$\sigma$ above the previous measurements. Both sources, however, are increased relative to other observations and HOPS-373 was already declining in submillimeter flux density. Together, this information suggests that the apparent increase is more likely a systematic flux density calibration error. The flux density measurements per-epoch are compiled in Table \ref{k_band_fluxes}.

While the overall flux density has not changed within the uncertainties, there is some qualitative evidence for small changes in the relative brightness of HOPS-373 SW vs. NE. The K-band continuum of the southwest source appears fainter than the NW source in 2015, while the ratios are reversed in 2021, suggesting that there may be increased emission related to the accretion burst, consistent with the change in Ka-band emission \citep{yoon2022}. Then in the final D- and C-configuration observations (2022 July - December), the K-band flux density of the SW source is no longer brighter than the NE source, indicating a relative change in the flux densities of the two sources, which could reflect the fading of the outburst. However, independent variability of the two sources could confound this examination of their relative flux density changes since we only know that they changed, not necessarily which one.

We attempted to quantitatively characterize the flux density ratio of the sources as a function of time, shown in the bottom panel of Figure \ref{HOPS-373-K-band-vs-time}. The ratio of the flux densities between the two sources is relatively constant over the monitoring period. The small excursions above or below a ratio of 1.0 are within the statistical uncertainties of a constant flux density ratio. Thus, despite the visually apparent changes in the images (Figure \ref{HOPS-373-K-band}), there is no significant detection of a relative change in flux density betweeen the sources in K-band.

Despite the apparent spatial filtering resulting is less recovered flux at the higher resolutions, no noteworthy structures are revealed toward either source. The sources appear compact and symmetric at the highest resolution 1.3~cm image generated with all data combined. This apparent lack of structure is consistent with the lack of structure in the slightly higher resolution 9~mm data from \citet{tobin2020}, but the compact dust continuum does appear marginally-resolved at 0.87~mm from observations presented in \cite{lee2023} at 0\farcs08 resolution.

\subsection{HOPS-373 Water Maser Emission}
\begin{figure}
\begin{center}
\includegraphics[scale=0.6]{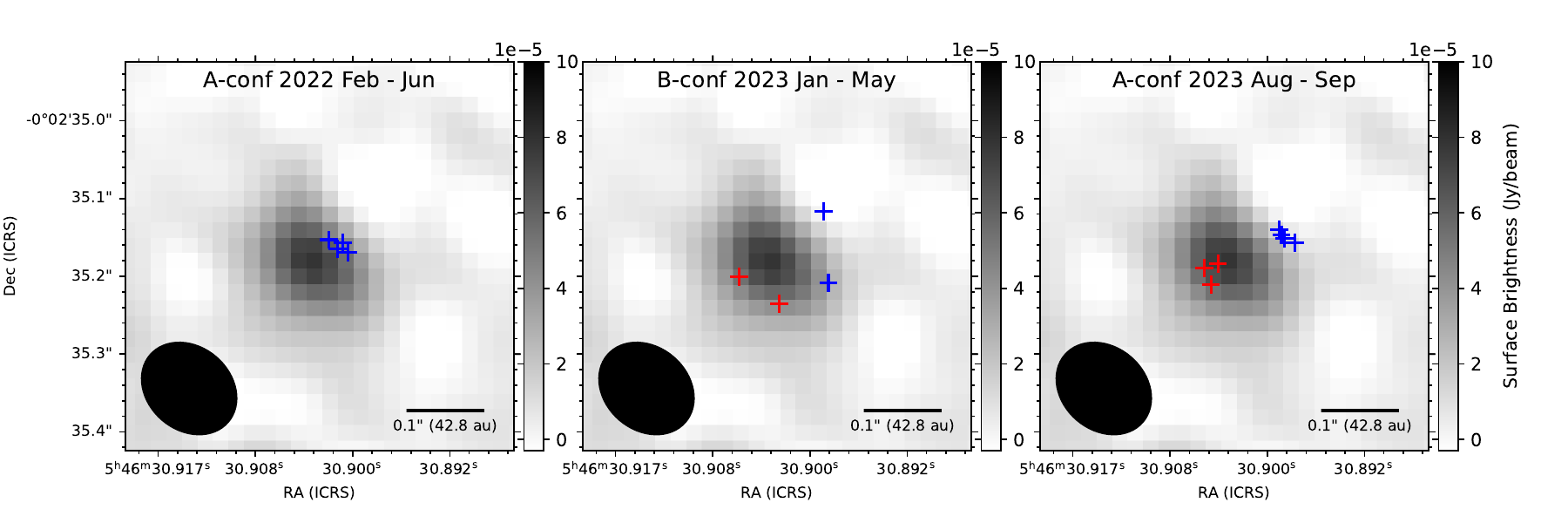}
\end{center}
\caption{Water maser positions detected during a particular visit to A and B configurations overlaid on the 1.3~cm image from A-configuration in 2022. We only overlay the masers on a single continuum image so that the motion relative to the same continuum position is more apparent and the masers are detected with much higher S/N than the continuum.}
\label{HOPS-373-K-band-zoom}
\end{figure}

We also monitored the water maser emission at 22.235 GHz toward HOPS-373, in conjunction with the 1.3~cm continuum. Water maser emission was confidently detected in 14 out of 23 observations (Table \ref{maser-properties}). The maser spot locations are overlaid on the 1.3~cm continuum data in Figures \ref{HOPS-373-K-band} and \ref{HOPS-373-K-band-zoom}, showing the typically close association of the maser spots and the southwest continuum source. The maser spots are not located directly coincident with the continuum peak, but tend to be offset slightly. The maser spots have the lowest positional accuracy in D-configuration, due to both the larger synthesized beam, but also the fact that the variable emission happened to be weakest during the D-configuration observations. 

We show the water maser spectra for each epoch of observation in Figure \ref{water-spectra} and \ref{water-spectra-b}, with Gaussian fits to the component(s) overlaid. It is clear from the spectra that the maser spots can change in flux, but they also change in velocity rapidly over time, with an observed velocity range of -15.2 to 21.3~\kms, well within the observed velocities of the CO outflow \citep{yoon2022,lee2024}. The systemic velocity of the source is $\sim$10.1~\kms\ \citep{lee2023} and thus there are maser components that are both red- and blue-shifted with respect to the source. Moreover, spots with red- and blue-shifted velocities can appear at the same time. The maser emission was not extremely strong, with the brightest maser having a peak of 0.2~Jy in one epoch. This is in contrast to the single-dish detection of a $>$10~Jy maser in the vicinity of HOPS-373 by \citet{haschick1983}. The masers observed by \citep{haschick1983} had velocities of $\sim$22~\kms, consistent with the most red-shifted components detected in our observations. \citet{haschick1983} used single-dish observations to search for and localize the masers, with a 72\arcsec\ beam, allowing for good localization toward HOPS-373 given that the nearest YSO is HOPS-403 $\sim$2\arcmin\ away. Archival VLA data (VLA project 15B-229) toward HOPS-403 do not detect a water maser, further evidence that the maser detected by \citet{haschick1983} was likely associated with HOPS-373.

\begin{figure}
\begin{center}
\includegraphics[scale=0.25]{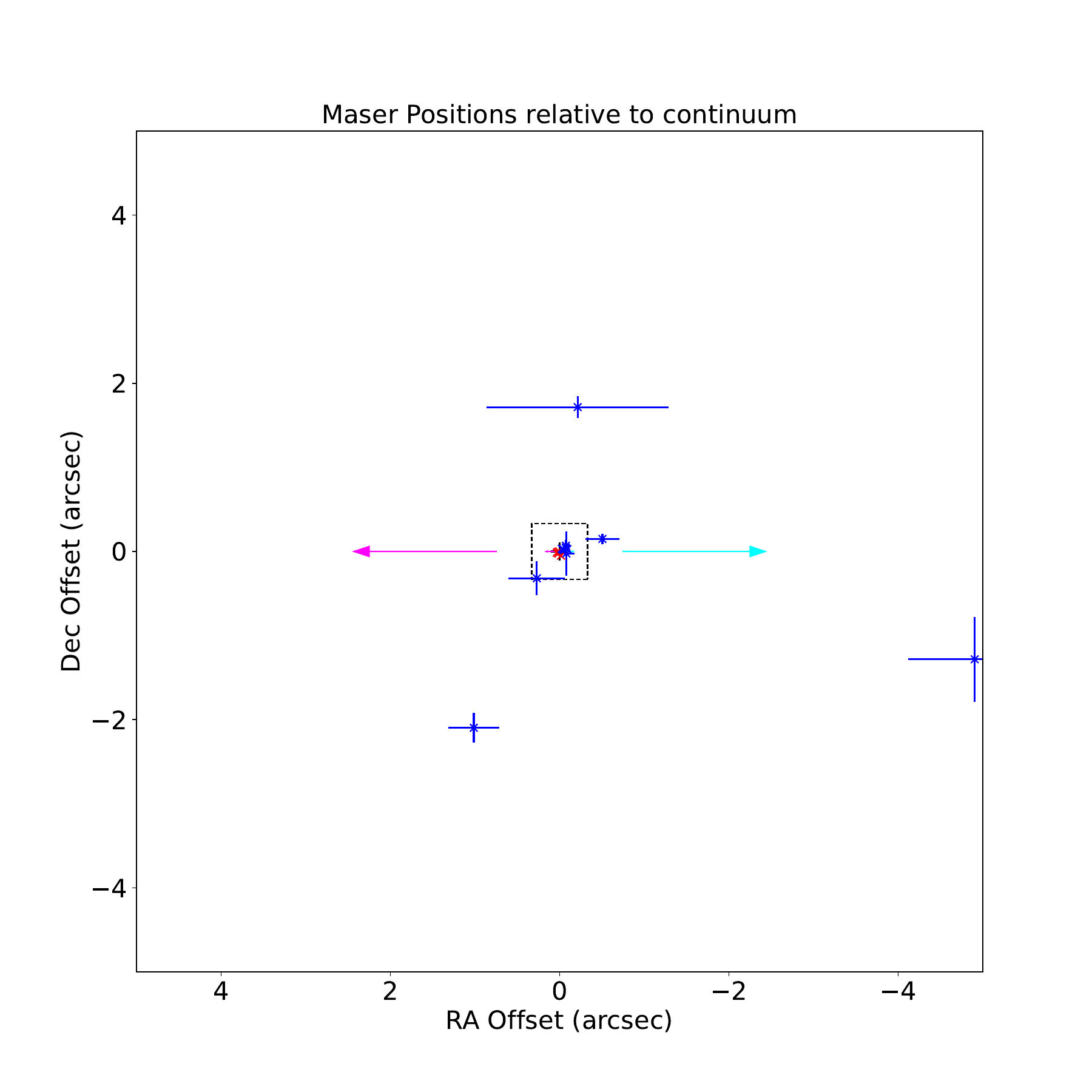}
\includegraphics[scale=0.25]{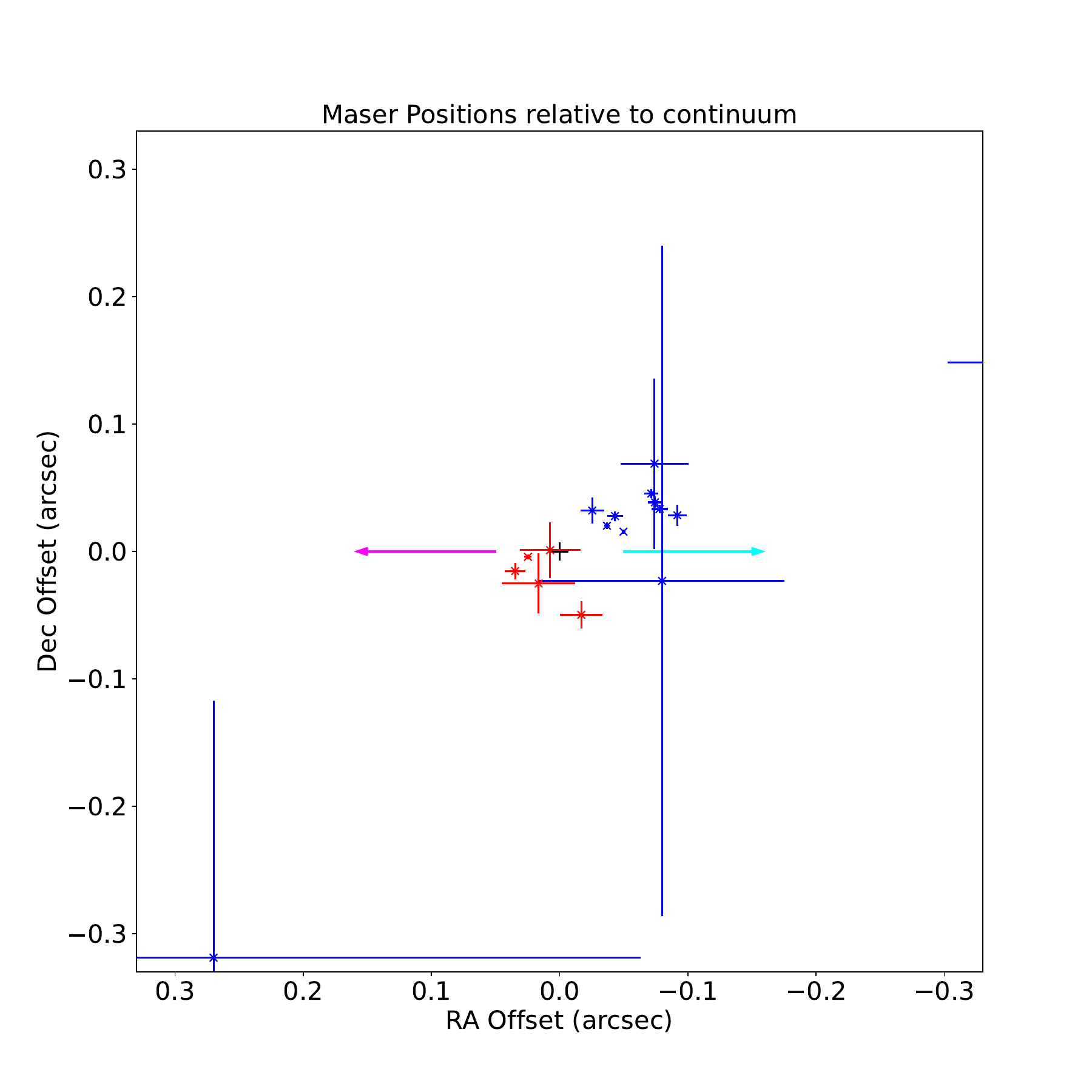}
\end{center}
\vspace{-5mm}

\caption{Water maser positions relative to the continuum source. The left panel shows a wider field of view that encompasses all detected maser positions, while the right panel shows a zoomed-in region toward the continuum source (denoted in the left panel as the dashed box). The systemic velocity of HOPS-373 is $\sim$10~\kms\ and the masers at blue-shifted relative velocities are plotted as blue points and those with red-shifted relative velocities are plotted as red points. The nearly east-west outflow is plotted as the cyan and magenta lines to denote the blue and red-shifted sides of the outflow, respectively. The continuum source is plotted as the black cross at (0,0). The maser positions outside the central $\pm$0\farcs33 are from D-configuration at lower angular resolution; no masers were detected in C-configuration observations.
}
\label{maser-positions}
\end{figure}

The blue- and red-shifted maser emission is not co-spatial. For the observations with high enough native resolution  (and S/N) there is a clear offset between the maser spots. We plot the maser positions in Figure \ref{maser-positions} relative to the continuum source and also draw the approximate outflow direction. The maser spots trace a line that is not exactly aligned with the outflow, but at an angle to its direction. However, the maser spots are more along the outflow than orthogonal to it. Furthermore, the velocity of the maser spots is opposite to the velocity gradient observed in complex organic molecules by \citet{lee2023}. Therefore we know the masers are not tracing a rotating ring as they can in some sources \citep[e.g.,][]{torrelles1998}. 

The maser spots could be moving, given that there is an increase in projected separation from the continuum position as a function of time, for the blue-shifted spots at least, as shown in Figure \ref{maser-motion}. However, the blue shifted spots earlier in time were preferentially at $\sim$3~\kms\, while the spots observed later in time were at $\sim$-10~\kms\ so we may not actually be tracing the same spots. This idea will be further discussed in Section 4.3.

\subsection{HOPS-373 Submillimeter to Radio Spectrum}

\begin{figure}
\begin{center}
\includegraphics[scale=0.6]{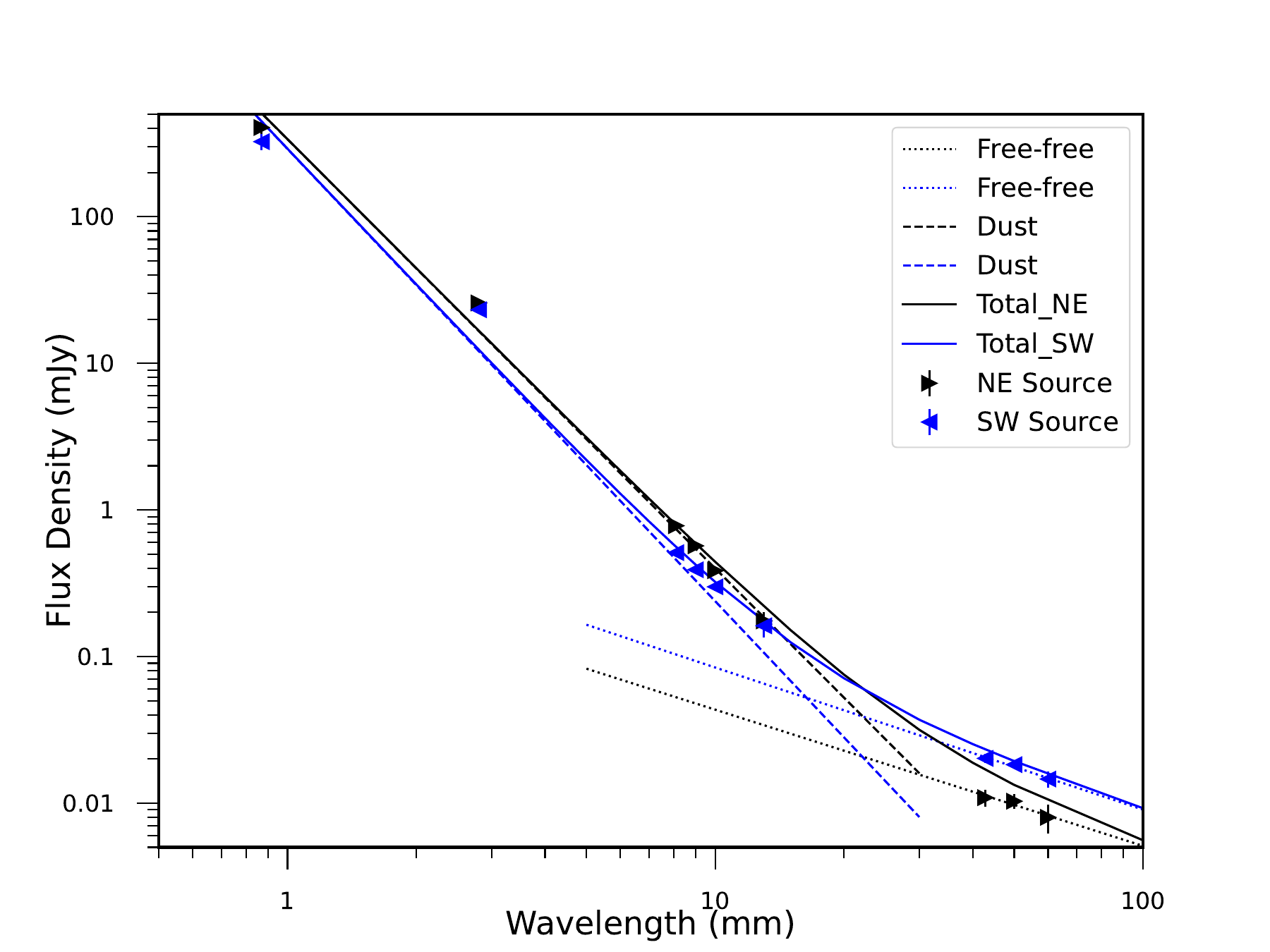}
\end{center}
\caption{Submillimeter to radio spectrum of the NE and SW components of HOPS-373. The dotted lines represent free-free emission fitted to only the three longest wavelength points and dashed lines represent dust emission. The individual flux density measurements corresponding to the points are given in Table \ref{radio-spectrum-data}. The total spectrum for NE clearly overshoots at the longest wavelengths, resulting from the shallower slope of the shorter wavelength data.
}
\label{radio-spectrum}
\end{figure}

Putting together data from the literature \citep{tobin2015,tobin2020,yoon2022,federman2023} and data presented here, we have assembled the spectra of HOPS-373 NE and SW from the submillimeter to radio in Figure \ref{radio-spectrum}, as measured by radio interferometers. We measured the flux densities toward HOPS-373 NE and SW by fitting Gaussians to the data from \citet{tobin2015,federman2023}; the original publications did not provide decomposed flux densities for the two sources. While the archival data from \citet{federman2023} had a beam 4\farcs99$\times$2\farcs90, they are marginally resolved and similar flux densities are found both with and without fixing the Gaussian parameters. The flux densities plotted in the in Figure \ref{radio-spectrum} are provided in Table \ref{radio-spectrum-data}.

What was clear from images shown earlier is that the SW source is clearly brighter at longer wavelengths than NE, but by 1.3~cm, again they are approximately the same in brightness. At 9~mm, however, both pre- and post-outburst NE is brighter than SW, and this was true both at $\sim$1\arcsec\ resolution \citep{yoon2022} and $\sim$0\farcs08 resolution \citep{tobin2020}. The brighter emission toward NE at 9~mm should not be taken to imply that NE was the source that underwent the burst because, at $\sim$0.87~mm with $\sim$0\farcs11 resolution, SW had a higher peak intensity than NE both pre- and post-outburst \citep{tobin2020,lee2023}, and that finding can be best reconciled if there is significant continuum opacity in NE that is suppressing the peak emission at 0.87~mm, relative to 9~mm, when observed at $\sim$0.1\arcsec\ resolution. The enhancement at 9~mm is unlikely to be from free-free emission because we know the free-free flux from NE is consistently quite low. The lower resolution 0.87~mm flux densities from \citet{federman2023} indicates that the NE source is slightly fainter than the SW source, but not significantly (blue/black points in Figure \ref{radio-spectrum}).

Furthermore, the overall flux density of dust emission toward NE was found to be higher in all observations, except those of \citet{lee2023} at $\sim$0\farcs08 resolution at 0.87~mm).
However, those data have the highest resolution of all datasets we considered, which can lead to more spatial filtering relative to the 0.87~mm data at $\sim$0\farcs11 resolution from \citet{tobin2020}. Thus, there may be two effects that impact the relative flux densities of NE and SW. First, NE could have high dust opacity, making its peak intensity fainter than SW at 0.87~mm. Second, lower opacity at longer wavelengths could then enable NE to be brighter than SW at $\sim$0\farcs08 at 9~mm \citep{tobin2020}. Finally, an extended component of the flux density toward NE enables its overall flux density to be higher than that of SW at wavelengths shorter than 1~cm at resolutions lower than 0\farcs1.

We fitted slopes to the free-free emission at 4 to 6~cm and used this fit to remove the expected contribution of free-free emission to the emission at 0.87~mm to 1.3~cm). We then fitted the free-free corrected 0.87~mm to 1.3~cm data to estimate the slope of the dust emission. There may be some uncertainties on the 0.87~mm and 2.8~mm flux densities given that they were observed in 2019 and 2013, respectively, but in both cases NE and SW have near equivalent flux densities.

The fitted free-free spectra can be described as a power-law in the form 
\begin{equation}
F_{\nu}~=~F_{\rm 30GHz}\nu^{\alpha}.
\end{equation}
NE and SW have $\alpha$ = 0.93~$\pm$~0.3 and 0.97~$\pm$~0.17, respectively, and $F_{\mathrm 30GHz,ff}$ = 0.043~$\pm$~0.03~mJy and 0.084~$\pm$~0.02~mJy, respectively. Then with the extrapolated free-free emission subtracted from the 8~mm to 1.3~cm emission, the dust emission is characterized for NE and SW by $\alpha$ = 2.9~$\pm$~0.14 and 3.1~$\pm$~0.2, respectively, and $F_{\mathrm 30GHz,dust}$ = 0.40~$\pm$0.013 and 0.24~$\pm$~0.015~mJy, respectively. The uncertainties are statistical only and do not take into account the 5-10\% flux calibration uncertainty. The spectral indices from the dust and free-free emission are consistent with typical values found for protostars at centimeter wavelengths \citep{tychoniec2018b}.


\section{Discussion}

The accretion and outflow processes are expected to be intimately linked \citep[e.g.,][]{frank2014}, with the outflow rate roughly expected to be $\sim$10\% of the accretion rate \citep{pelletier1992,blandford1982}. Thus, for an accretion burst, like the one observed toward HOPS-373 by \citet{yoon2022}, we would naively expect that the increased accretion rate should also temporarily increase the outflow rate. Furthermore, the $\sim$25\% increase in submillimeter flux measured by the JCMT is estimated to indicate $\sim$4$\times$ a luminosity increase given that the submillimeter dust emission is primarily informing us of the increased dust temperature from the higher luminosity \citep{johnstone2013,contreraspena2020}. The only way to increase the overall luminosity of a young star in a short period of time is to increase the accretion rate from the disk to the protostar, and accretion luminosity is directly proportional to the accretion rate.

A theoretical model describing radio continuum emission from collimated, ionized winds was described by \citet{reynolds1986}. Within the context of this model, the ionized outflow rate is directly proportional to the radio continuum flux density. Thus, an increased accretion rate of $\sim$4$\times$ should lead to the same increase in the outflow rate, for the period of high accretion. The timing of the accretion bursts was such that there was a first burst in late 2019, with a rapid fading, followed by another burst in late 2020. The second burst resulted in a higher submillimeter flux for about a year and has been fading, but had not returned to the pre-2019 flux density levels by the end of JCMT monitoring in 2024 (Figure \ref{C-band-vs-time-wJCMT}). Thus, the first burst may have resulted in a brief period of higher outflow rate, while the second burst should have resulted in a sustained period of higher outflow rate.

\subsection{On the Lack of Increased Free-Free Emission Toward HOPS-373}

The outbursts observed in the submillimeter and near-infrared for HOPS-373 lead us to naively expect that the outflow rate of the protostars should also correspondingly increase. However, the lack of a statistically significant change ($<$66\% or lower if the flux density changes between configurations are ignored) in the 5~cm flux density throughout the entire period of monitoring clearly shows that the observed luminosity burst did not result in a free-free flux density change. Furthermore, even going back to 2015, there was not statistically significant change in the 5~cm flux density. If the free-free emission levels tracked the changes in accretion, we would have expected an overall increase in 2021 relative to 2015 and then perhaps a fading from late 2022 through 2023 as the luminosity decreased. Neither of these scenarios occurred and we do not think the apparent decrease in 5~cm flux density at the end of monitoring was the result of decreasing intrinsic flux density, but rather filtered-out emission from the higher resolution configuration.

The lack of a response to the increased luminosity is not unique to HOPS-373. \citet{galvanmadrid2015} examined archival data toward HOPS-383, a Class 0 protostar that had a luminosity increase of $\sim$35$\times$ \citep{safron2015}. They examined archival data from the historical VLA pre-outburst and VLA data post-outburst, not finding a significant increase in flux density at 5~cm. Then \citet{forbrich2015} examined the source Per-emb-28 (LLRL 54361), where periodic accretion events are witnessed due to binary modulated pulsed accretion \citep{muzerolle2013} with a period of $\sim$29 days. There was not a significant difference in the C-band flux density measured at the peak of accretion luminosity or between accretion events. The change in luminosity (and hence the accretion rate) with each event was a factor of $\sim$10.

The authors of the aforementioned works discussed that the free-free emission must be operating on a different timescale relative to the accretion events, given that the accretion happens very close to the star while the free-free emission may be coming from radii of up to $\sim$~100~au, and jets knots are sometimes spatially resolved \cite[e.g., Section 3.4, ][]{anglada2018}. For example, if an outflow knot is ejected in association with an accretion event it may propagate at a velocity of $\sim$100~\kms\ \citep{frank2014,anglada2018}, which translates to 20~au~yr$^{-1}$. Thus, with this propagation velocity, an ejected shock knot might be hidden behind optically-thick free-free emission, or emission from the knot itself could be optically thick.

The other important aspect of the \citet{reynolds1986} model to highlight that it is the ionized mass outflow rate that is correlated with the radio flux. Thus, if there is an increase in the outflow rate and it remains mostly neutral or molecular \citep{ray2023}, it will not contribute to the free-free flux. \citet{hartmann2016} further pointed out the winds from FU Ori systems (a class of outbursting young stars) tend to be neutral, which might provide a clue to the lack of increase in ionized emission. Furthermore, the amount of free-free emission could also depend on the outflow velocity in the \citet{reynolds1986} model. Therefore, if it is possible to increase the outflow rate, but also slow down the outflow velocity, then a change in flux may not be possible. This could be understood in terms of momentum conservation, if the wind or outflow is launched with constant momentum, then an increase in the outflowing mass will need to have a lower velocity.

The monitoring we conducted at 5~cm was some of the most extensive toward a particular group of protostars. Thus, the lack of a response to the accretion burst provides important constraints for models of outflow ejection following accretion events, suggesting that the response from free-free emission may be significantly delayed (or non-existent) from the occurence of an accretion burst. In particular, if a shocked jet is responsible for the free-free emission, a knot may need to travel many tens of au to emerge from the optically thick region of free-free emission for its emission to become detectable, which could ultimately be related to molecular knots with observed or presumed proper motions and may be correlated with past outbursts \citep{plunkett2015, reipurth1989, ray2023, kim2024}. However, the addition of the 2026 February data extends the time baseline to $\sim$5~yr since the burst began and a knot could have travelled $\sim$100~au (0\farcs25) and should have become detectable if present.

\subsection{Ambiguity of the Burst Detection in 1.3~cm Continuum}

The 1.3~cm continuum emission is expected to have a large ($>$50\%, see Section 3.7) contribution from thermal dust emission, thus its overall flux density is should have some direct response to the heating from the outburst event, even if the free-free emission remained constant (as characterized at 5~cm). The fractional change in flux density of the burst observed with the JCMT at its peak was $\sim$0.3 \citet{yoon2022}. Therefore, if the flux density from dust emission (expected F$_{\mathrm dust}$ $\sim$ 0.075-0.1~mJy; Section 3.7) observed at 1.3~cm changed by a similar amount, the increase in flux density would be just $\sim$20-30~\microjy, which comparable to the statistical uncertainties of the 1.3~cm flux densities (Table \ref{k_band_fluxes}). This means that any increase in 1.3~cm dust emission from the burst is not likely to result in a statistically significant change in the observed flux density. Assuming that the NE source remains at a constant flux density, the expected fractional change in relative flux densities is only $\sim$10\% for a burst that fractionally increased the dust emission by at most $\sim$30\%. Thus, with the observations presented here, we would have only been able to obtain a positive detection of variation between sources if both the dust and free-free emission varied at the 30\% level or greater.


\subsection{Increased Free-free Emission in HOPS-321}

While HOPS-373 showed no increase in its free-free emission, HOPS-321 showed an increase in free-free emission by $\sim$1.5$\times$ during the course of our monitoring. At the same time, the near-infrared flux density from HOPS-321 monitored by NEOWISE has remained relatively steady, as has the submillimeter flux density measured by the JCMT. Thus, the free-free emission of this source increased rapidly without any corresponding increase in emission at other wavelengths throughout the monitoring period. HOPS-321 is Class I protostar with a bolometric luminosity and temperature of 3.7~\lsun\ and 78.6~K \citep{furlan2016, tobin2020}, and the spectral index of emission ($\alpha_{pre-increase}$ = 0.82 and $\alpha_{post-increase}$ = 0.66) indicated that it is most likely free-free emission and very unlikely to be gyrosynchrotron, which could have a negative spectral index.

Thus, in the absence of an associated increase in the accretion luminosity and gyrosynchrotron, something else must result in the flux density increase of the free-free jet. As mentioned earlier, the outflow velocity can also influence the amount of radio emission. Thus, if the outflow velocity could increase without a corresponding change in luminosity, it could be possible for the free-free emission to change and possibly not show a response at shorter wavelengths (i.e., X-rays). Alternatively,
if the burst that resulted in the increased free-free emission was short-lived, it might not have been detected by NEOWISE nor the submillimeter monitoring because of the 6 month gaps in NEOWISE monitoring and if the start of the 5~cm flux density increase was at the start of D-configuration (2023 July 23 or 59783 MJD), then the burst could have occurred during the $\sim$4 month gap in JCMT monitoring corresponding to Orion being above the horizon during daytime.

The review by \citet{anglada2018} pointed out the case of B335, an isolated, low-luminosity protostar, where the free-free flux density has had large amplitude changes over the years. However, there were no constraints on whether the overall luminosity of B335 was varying or if it was just the free-free emission. Measurements by NEOWISE in the past decade, however have shown that B335 had an increase in luminosity and subsequently faded \citep{kim2024}. Thus, it is possible that the free-free emission responds on a much longer and indeterminate time scale relative to physical processes like accretion, but we also cannot rule-out a short-term burst that was missed due to gaps in monitoring cadence.

\subsection{Maser Activity in Relation to Outburst}

Despite the free-free emission having no evidence for variation toward HOPS-373, we found large changes in the water maser flux densities toward the variable source HOPS-373 SW. The maser spots appeared and disappeared somewhat randomly around source, but some of the strongest maser spots were coincidentally observed when the VLA was at its highest resolution in A-configuration. These masers are, however, much fainter and at different velocities than the single-dish detection by \citep{haschick1983}. 

Maser spots trace a line at a $\sim$30\degr\ angle with respect to the main molecular outflow, as shown in Figure \ref{maser-positions}. Moreover, \citet{lee2024} showed evidence that there appear to be two outflows from HOPS-373 SW, or that the outflow has changed direction. One of the outflows is mostly in the East-West direction, while the other deviates slightly from East-West with a position angle of $\sim$-80\degr\ East of North. This is not at as large of an angle as the maser spots, but the blue- and red-shifted maser spots are on the same sides as the blue and red-shifted molecular outflows.

\begin{figure}
\begin{center}
\includegraphics[scale=0.6]{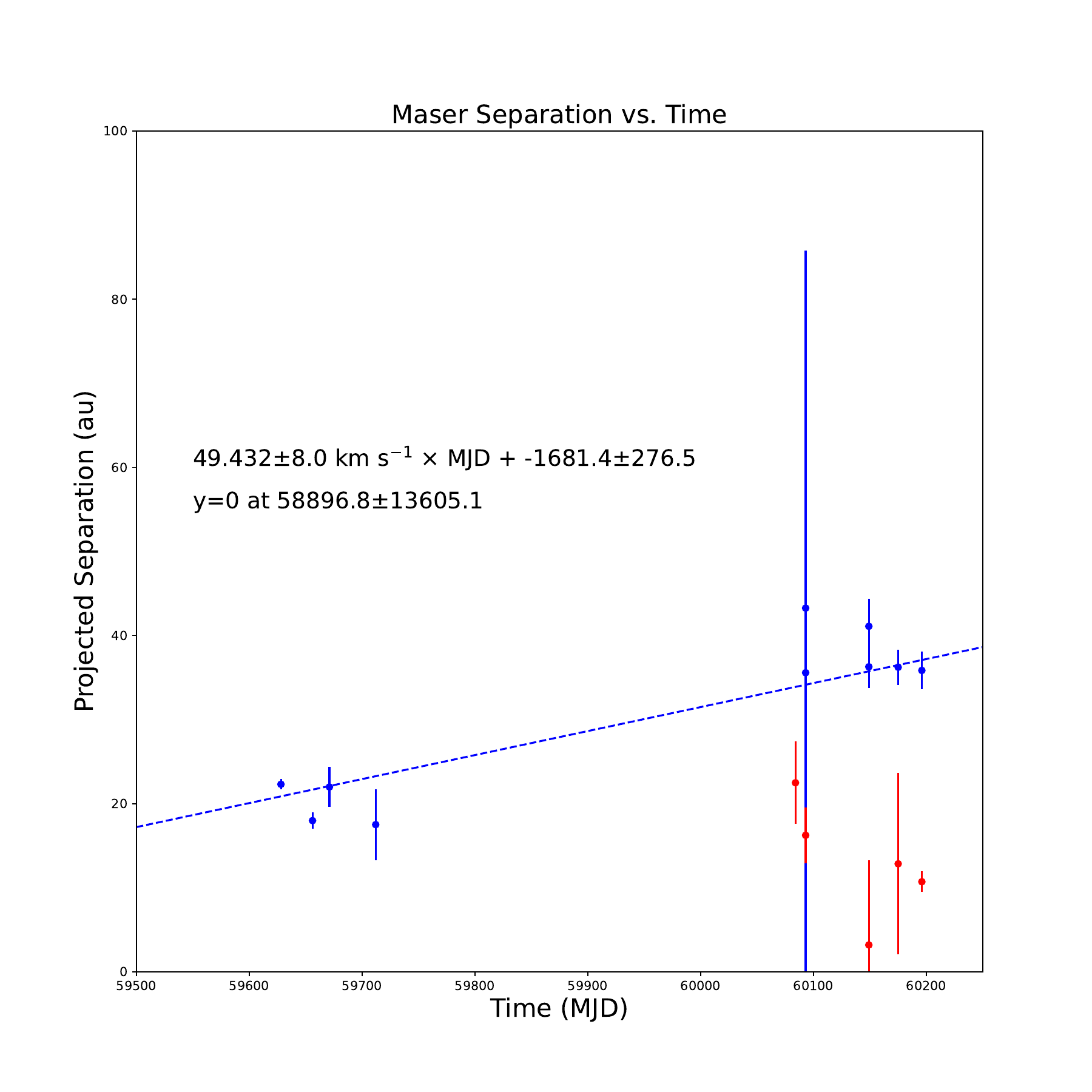}
\end{center}
\vspace{-15mm}

\caption{Projected separation of water maser positions relative to the continuum source versus the MJD the masers were detected. The blue-shifted masers detected between 59600 to 59750 and 60100 to 60200 are fitted with a line given that these masers resemble outward motion. Their origin at r=0 could be at $\sim$59000, but with very large uncertainty. Not all masers appear on this plot, particularly those with positions $>$0\farcs25 from the continuum source.
}
\label{maser-motion}
\end{figure}

Furthermore, the blue-shifted maser spots measured in 2022 relative to 2023 show that they are further from the continuum source and a linear fit suggests that the could be traveling at $\sim$50~\kms\ in projected velocity (Figure \ref{maser-motion}). The red-shifted spots, however, show a different behavior and are located closer to the continuum source with time, but no red-shifted masers were detected in 2022. Thus, these maser spots could be associated with shocks within a precessing outflow, and \citet{lee2024} suggests that the molecular outflow could be precessing. 

Thus, the maser emission and the apparent shift in position with time could result from a shockwave propagating through the outflow cavity. Then as the shock passes a given region, it could create the conditions for water maser emission, and then after the shock passes the conditions to produce maser activity are no longer present and the emission fades. Therefore, it could be possible that the masers appear on the 'crests' of a shockwave propagating from the protostar. The linear fit to the blue-shifted masers (taken at face value) in Figure \ref{maser-motion} indicates that they could have originated from the continuum source at MJD $\sim$59000, approximately the time of the second outburst. We caution, however, that the uncertainties on the origin time are quite large, so this scenario is quite tentative and subsequent maser observations would be useful to determine if they are still (or actually) propagating outward.

\section{Summary and Conclusions}

We have presented the results of a multi-year monitoring campaign toward the protostar HOPS-373 in continuum emission at 5~cm and 1.3~cm, as well as water maser emission. We also present the monitoring results for all sources detected within the 5~cm primary beam, with a focus on the known protostars. 

We did not detect variability toward HOPS-373 NE nor SW between 2015 and 2026 in at 5~cm, which encompasses the pre-burst, near the burst peak in 2021, the decline from 2021 to the end of monthly 5~cm monitoring in 2023, and the final epoch in 2026. HOPS-321, however, was found to vary, where the 5~cm flux density increased by about a factor of 1.5$\times$ in $\sim$ 2022 September and remained at the higher flux density through 2026 February. Many other YSOs in the 5~cm field of view exhibited stochastic variability, some of which had very high amplitude, which is likely attributable to gyrosynchrotron radiation.

The lack of change in the 5~cm flux density corresponding to the submillimeter outburst in HOPS-373 indicates that there is not a clear link between modest luminosity bursts and free-free emission. The luminosity bursts are expected to reflect an increase in accretion rate, which are expected to also result in a corresponding increase in outflow rate. If accretion and outflow are linked, then the free-free emission may only respond on a significantly different timescale relative to any accretion burst or the increase in the outflow rate may be mostly in neutral or molecular gas and have less of an impact on the ionized outflow rate. Furthermore, the increase in free-free emission for HOPS-321, with no corresponding evidence of an accretion burst in the submillimeter or NEOWISE, indicates that free-free emission may vary independent of any change in luminosity, at least for the timescales of the JCMT Transient Survey, NEOWISE, and the VLA monitoring.

The maser emission from HOPS-373 is extremely variable in its flux and there are multiple distinct velocity components. There is not a clear relationship between the outburst and the maser emission, but the maser could be tracing newly generated shocks in the outflow that are propagating away from the protostar.

\begin{acknowledgements}
We acknowledge the assistance from the VLA Data Analysts in performing the Quality Assessment on the data and flagging as necessary. We thank NRAO staff member P. Beaklini for his efforts to devise the spectral setup that we used for the final A-configuration C-band data that avoids receiver gain compression from RFI. D.J. is supported by NRC Canada and by an NSERC Discovery Grant. G.J.H. is supported by the National Key R\&D program of China 2022YFA1603102 from the Ministry of Science and Technology (MOST) of China. J.-E. Lee was supported by the National Research Foundation of Korea (NRF) grant funded by the Korea government (MSIT) (grant numbers RS-2024-00416859 and RS-2026-25490557). The National Radio Astronomy Observatory and Green Bank Observatories are facilities of the U.S. National Science Foundation operated under cooperative agreement by Associated Universities, Inc.
\end{acknowledgements}
 \facility{VLA}
\software{Astropy \citep[http://www.astropy.org; ][]{astropy2013,astropy2018,astropy2022}, 
APLpy \citep[http://aplpy.github.com; ][]{aplpy}, CASA \citep[http://casa.nrao.edu; ][]{mcmullin2007}}

\appendix

\section{5~cm Continuum of Other Sources}

The full list of continuum sources detected in the B-configuration observations is presented in Table \ref{c-band-catalog}; this catalog is generated from the B-configuration observations observed in 2023. We regard B-configuration observations as our most robust dataset for catalog generation because many sources that were blended at lower resolution are resolved, but the emission is not so well-resolved that it leads to non-detections. For each detected source, we attempted to match it to known sources using Simbad and the cataloged HOPS protostars in the region. Many sources have no known counterparts and many are faint. The faint sources without known counterparts at other wavelengths are likely to be background extragalactic sources \citep{anglada1998}.

\begin{figure}
\begin{center}
\includegraphics[scale=0.8]{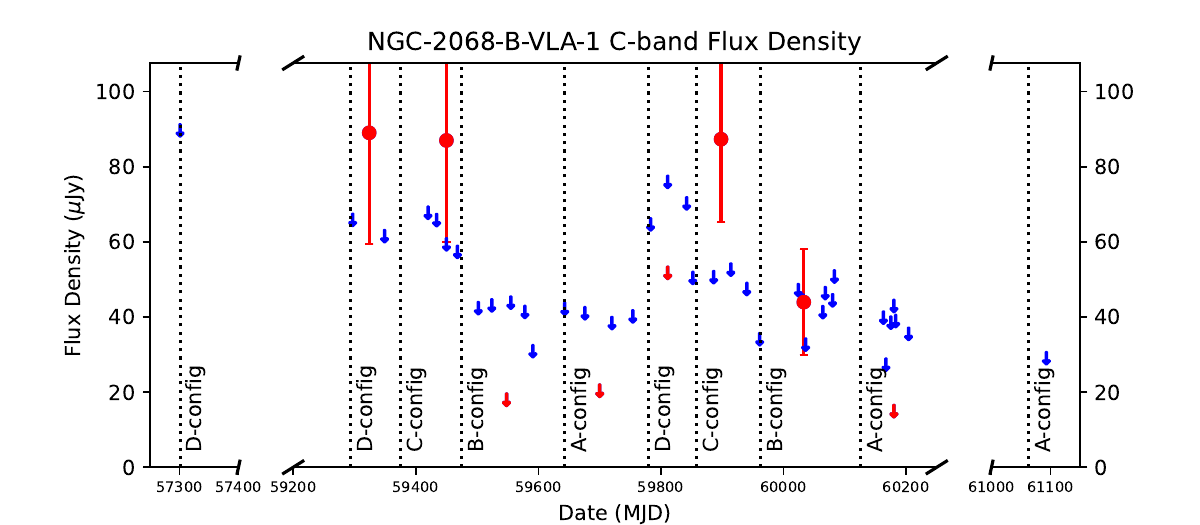}
\end{center}
\caption{Plots of 5~cm flux density versus time for each source detected. The vertical lines denote changes in the VLA configuration. The blue points denote the flux density measurements from individual observations, while the downward blue arrows denote upper limits. The red points denote measurements using all data from a particular configuration combined and the downward red arrows denote upper limits from the combined images. The complete figure set (96 plots) is available in the online journal.}
\label{all-source-time-plots}
\end{figure}

We used the B-configuration source catalog to measure the flux densities versus time for each detected source. The individual flux density measurements (and upper limits) as a function of time are provided in Table \ref{per-epoch-fluxes}. The plots of flux density versus time are shown for all sources in the catalog in Figure \ref{all-source-time-plots}. If a source was blended with others in the lower-resolution data from D or C-configuration, we simply plot the flux density of the composite source along with the resolved flux densities from A and B configuration. Compared to A-configuration, there did not appear to be a case where a single source detected in B-configuration was clearly resolved into two or more sources in A-configuration.

Those sources where we manually measured the flux densities using Gaussian fits in the main text (Table \ref{per-epoch-imfit-fluxes}), tend to have larger uncertainties on their individual flux densities from the automated measurements, but there is good consistency between the measured values and behavior of the light curves. However, in cases like HOPS-373, both sources are not always detected in blind source searches due to their faintness, but we were able to more reliably measure the flux densities using \textit{a priori} knowledge of the source positions.

Some of the detected sources have highly variable 5~cm continuum emission, notably VLA-16, 37, 48, 88 and we show those light curves (Figure \ref{all-source-time-plots}). The 5~cm flux densities can range from non-detections to hundreds of $\mu$Jy to even mJy flux densities. These most highly variable sources are known to be young stars that are no longer embedded within envelopes. Previous surveys of young stars at 5~cm have identified such variable systems as emitters of gyrosynchrotron radiation, presumably from magnetic reconnection events in their stellar coronae \citep[e.g.,][]{dzib2013}.

\begin{small}
\bibliographystyle{aasjournalv7}
\bibliography{ms}
\end{small}

\clearpage

\begin{figure}
\begin{center}
\includegraphics[scale=0.75]{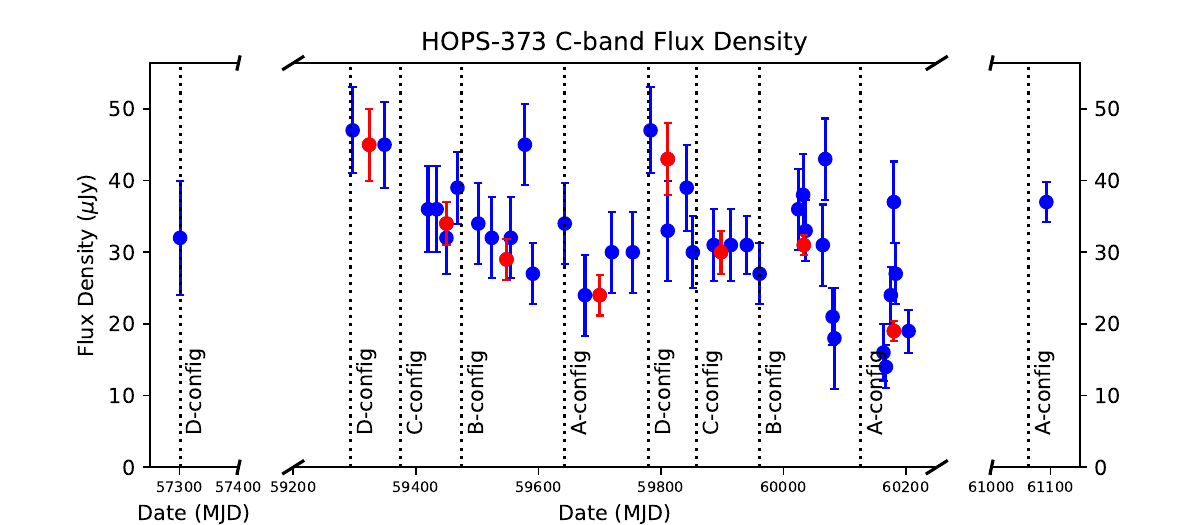}
\includegraphics[scale=0.75]{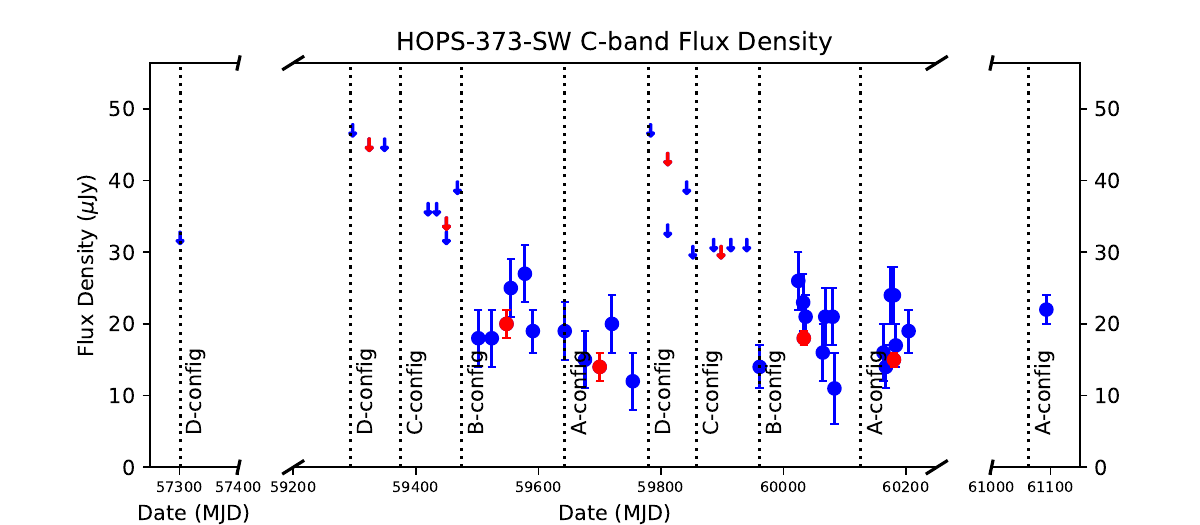}
\includegraphics[scale=0.75]{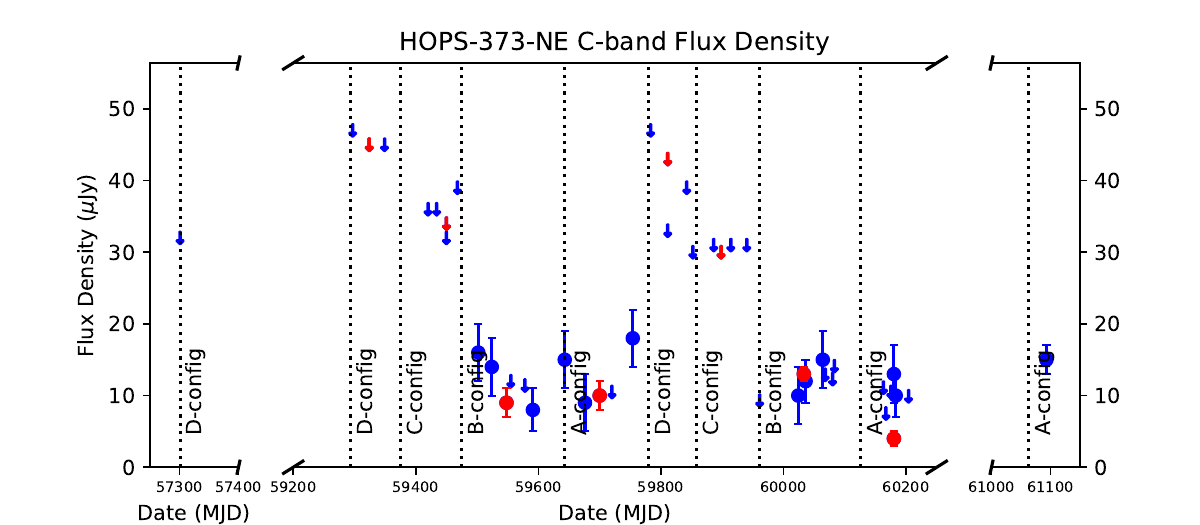}
\end{center}
\caption{Flux density measurements for HOPS-373 (total flux) and HOPS-373 SW and NE from all the epochs of observation. The blue points are measured from the individual observations, while the red points are measured using an image generated for all the data in a particular configuration. HOPS-373 has a slight decline in flux density that occurs with observations in higher resolution configurations, as such, we do not think there is evidence for significant variation in its 5~cm flux density. The flux densities for SW and NE are all upper limits for D and C-configurations because they are not resolved from each other and we simply use the total flux density of HOPS-373 as an upper limit. Even the final monitoring point in 2026 February is consistent with the last measurements in 2023.
}
\label{C-band-vs-time-H373}
\end{figure}

\begin{figure}
\begin{center}
\includegraphics[scale=0.65]{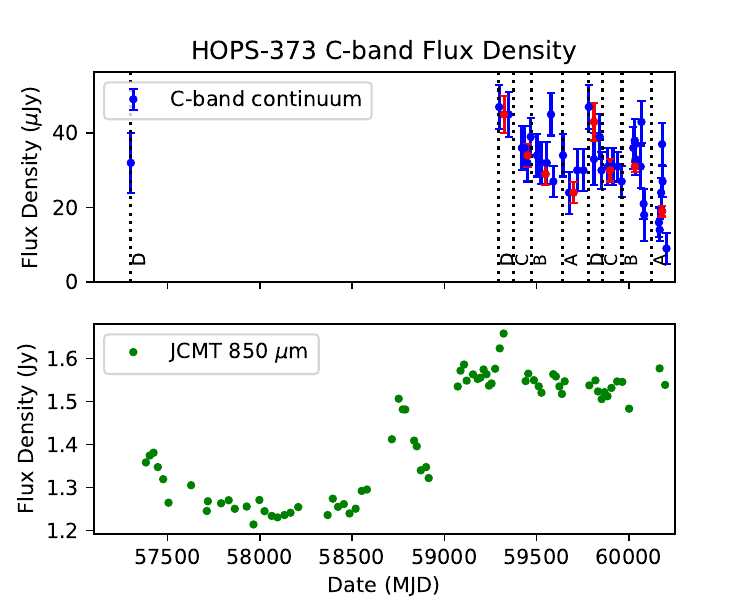}
\includegraphics[scale=0.65]{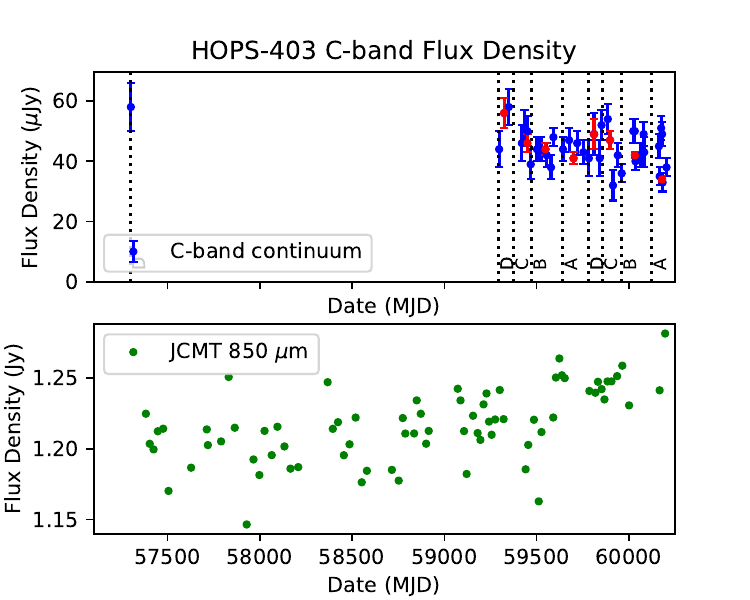}
\includegraphics[scale=0.65]{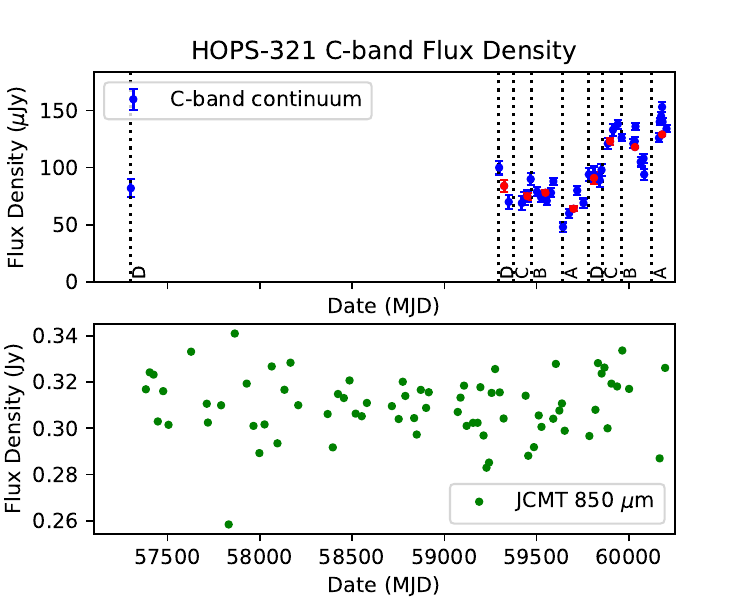}
\end{center}
\caption{Flux density measurements for HOPS-373, HOPS-403, and HOPS-321 at 5~cm (top panels) and the JCMT 0.85~mm flux densities for the full monitoring period of the JCMT transient survey (bottom panels). The 5~cm points are plotted in the same manner as in Figure \ref{C-band-vs-time-H373} and Figure \ref{C-band-vs-time-others}.
}
\label{C-band-vs-time-wJCMT}
\end{figure}

\clearpage

\begin{figure}
\begin{center}
\includegraphics[scale=0.6]{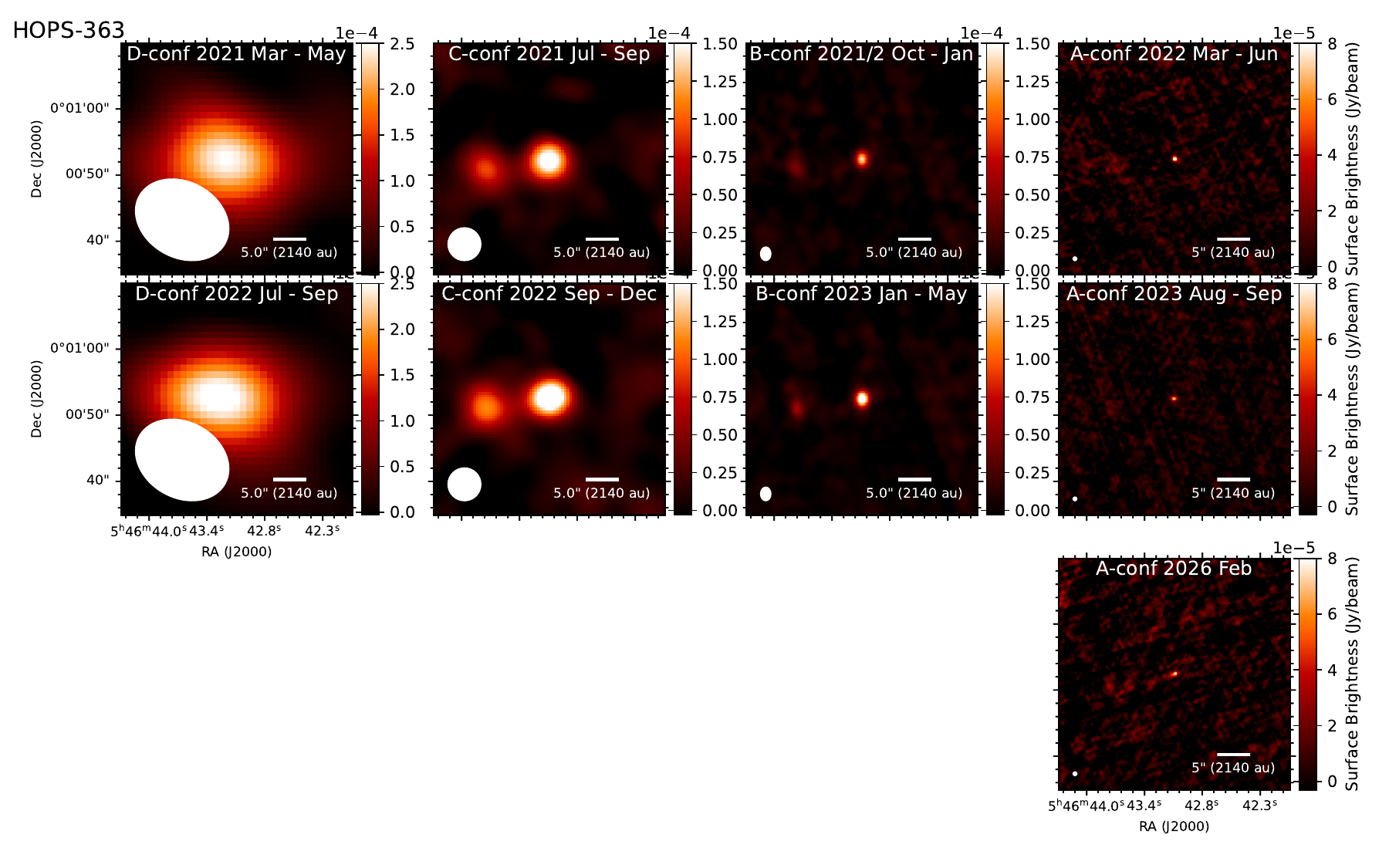}
\end{center}
\caption{Same as Figure \ref{HOPS-373-C-band} but for HOPS-363 and all panels show the same field of view. The D-configuration images clearly appear resolved in the east-west direction and the C- and B-configuration images clearly show a source resolved to the east. This is in the direction of the red-shifted outflow from HOPS-363 and the source could be associated with the outflow.
}
\label{HOPS-363-C-band}
\end{figure}

\begin{figure}
\begin{center}
\includegraphics[scale=0.8]{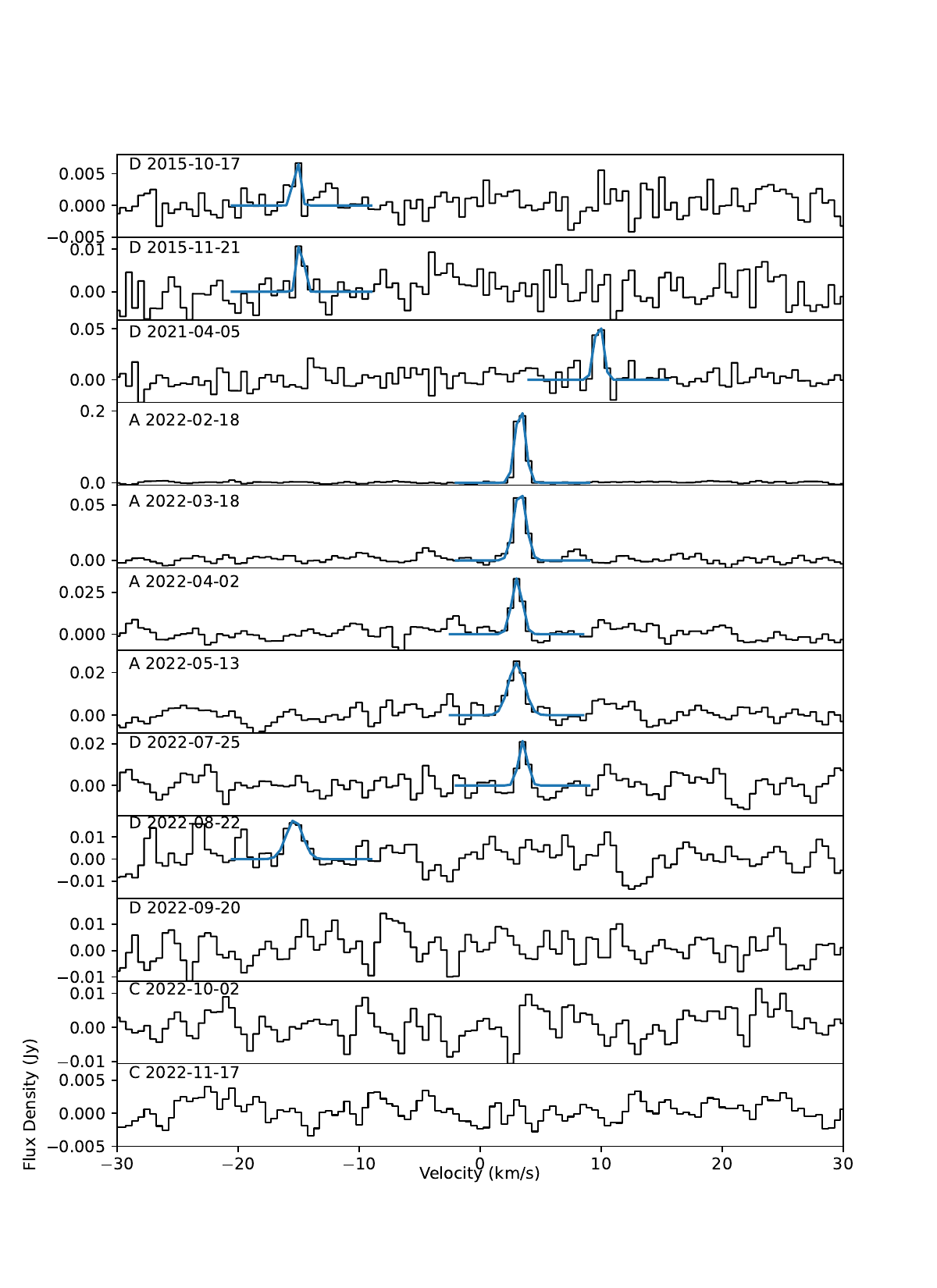}
\end{center}
\vspace{-20mm}
\renewcommand{\thefigure}{\arabic{figure}a}
\caption{Spectra around the water maser line at 22.235 GHz for each epoch of observation (black line). Spectra with a detection have a Gaussian fit overlaid (blue or orange). The spectra for 2023-05-29 and 2023-07-24 have two components fitted to their feature around -10~\kms. Full detail of the masers are provided in Table \ref{maser-properties}.
}
\label{water-spectra}
\end{figure}

\begin{figure}
\begin{center}
\includegraphics[scale=0.8]{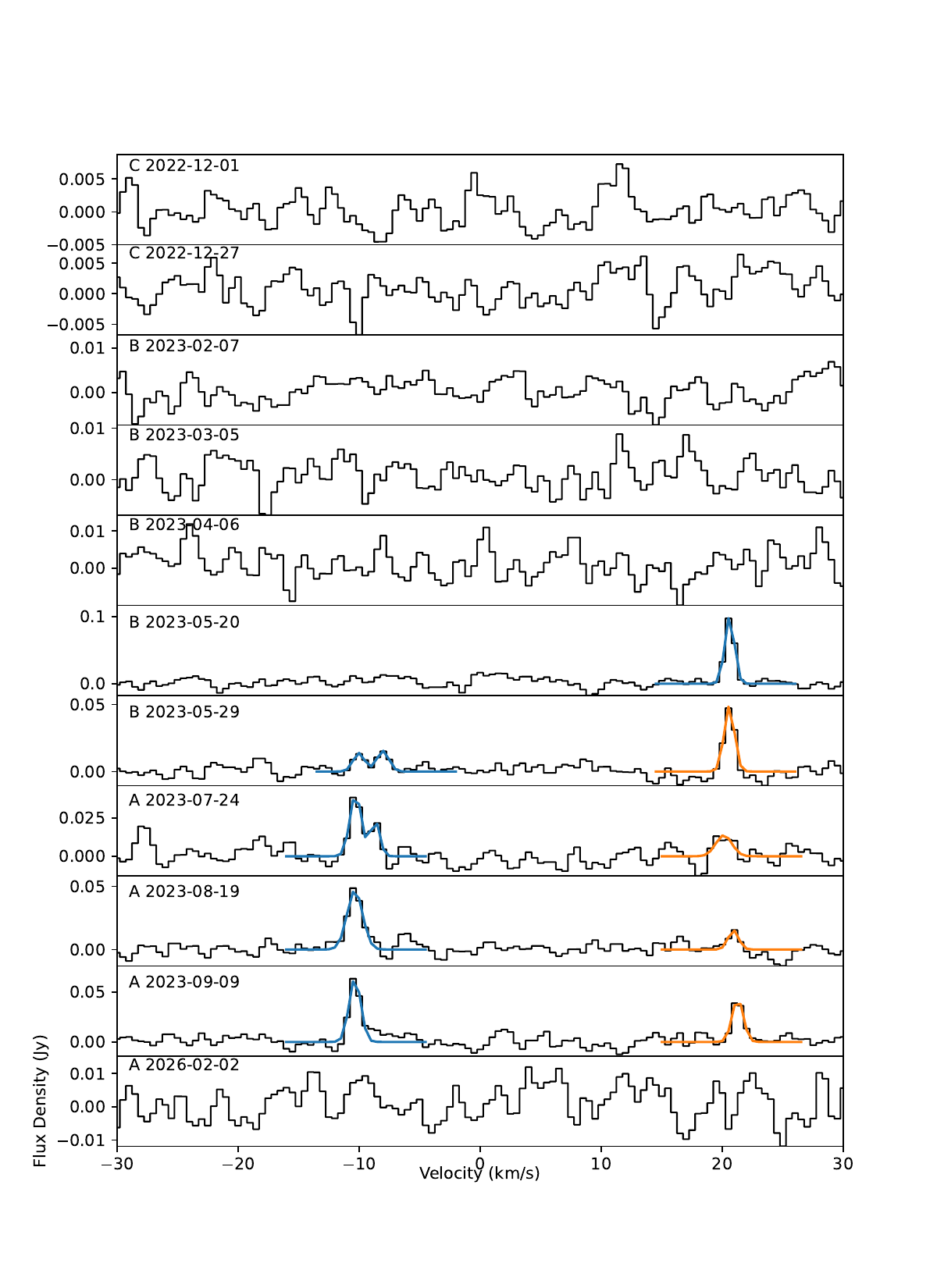}
\end{center}
\addtocounter{figure}{-1}
\renewcommand{\thefigure}{\arabic{figure}b}
\vspace{-20mm}

\caption{
}
\label{water-spectra-b}

\end{figure}

\clearpage
\figsetstart
\figsetnum{16}
\figsettitle{Time Series Plots for All Sources}
\figsetgrpstart
\figsetgrpnum{16.1}
\figsetgrptitle{NGC-2068-B-VLA-1}
\figsetplot{per_source_tables/NGC-2068-B-VLA-1-Flux-vs-time.pdf}
\figsetgrpnote{Time Series Plot for NGC-2068-B-VLA-1}
\figsetgrpend
\figsetend

\clearpage

\begin{deluxetable}{llcclllllll}
\tabletypesize{\scriptsize}
\tablewidth{0pt}
\tablecaption{VLA C-band Observation Log}

\tablehead{\colhead{Proposal Code} & \colhead{Date} & \colhead{Date} & \colhead{EBs} & \colhead{Config.} & \colhead{Duration} & \colhead{Samplers} & \colhead{Calibrators} & \colhead{Scale Factor} & \colhead{Selfcal?}\\
 & \colhead{(YYYY-MM-DD)} &  \colhead{(MJD)} &        &      & \colhead{(hr)}   &  &\colhead{(Complex Gain, Bandpass, Flux)} & \colhead{} &
}
\startdata
15A-369 & 2015-10-06 & 57301 & 2 & A$\rightarrow$D\tablenotemark{a} & 1.0 & 3-bit & J0541--0541, 3C147, 3C147 & \nodata & Yes\\
21A-409 & 2021-03-24 & 59297 & 1 & D & 1.2 & 3-bit & J0552+0313, 3C147, 3C147 & 0.950 & No\\
21A-423 & 2021-05-15 & 59349 & 1 & D & 1.3 & 3-bit & J0552+0313, 3C147, 3C147 & 0.999 & No\\
21A-423 & 2021-07-25 & 59420 & 1 & C & 1.3 & 3-bit & J0552+0313, 3C147, 3C147 & 1.035 & No\\
21A-423 & 2021-08-08 & 59434 & 1 & C & 1.3 & 3-bit & J0552+0313, 3C147, 3C147 & 1.031 & No\\
21A-423 & 2021-08-24 & 59450 & 1 & C & 1.3 & 3-bit & J0552+0313, 3C147, 3C147 & 1.030 & No\\
21A-423 & 2021-09-11 & 59468 & 1 & C & 1.3 & 3-bit & J0552+0313, 3C147, 3C147 & 1.087 & No\\
21A-423 & 2021-10-15 & 59502 & 1 & B & 1.3 & 3-bit & J0552+0313, 3C147, 3C147 & 1.075 & No\\
21A-423 & 2021-11-06 & 59524 & 1 & B & 1.3 & 3-bit & J0552+0313, 3C147, 3C147 & 1.046 & No\\
21A-423 & 2021-12-07 & 59555 & 1 & B & 1.3 & 3-bit & J0552+0313, 3C147, 3C147 & 1.009 & No\\
21A-423 & 2021-12-30 & 59578 & 1 & B & 1.3 & 3-bit & J0552+0313, 3C147, 3C147 & 0.990 & No\\
21A-423 & 2022-01-12 & 59591 & 1 & B & 1.3 & 3-bit & J0552+0313, 3C147, 3C147 & 1.040 & No\\
22A-268 & 2022-03-05 & 59643 & 1 & A & 1.3 & 3-bit & J0552+0313, 3C147, 3C147 & 1.103 & No\\
22A-268 & 2022-04-07 & 59676 & 1 & A & 1.3 & 3-bit & J0552+0313, 3C147, 3C147 & 1.012 & No\\
22A-268 & 2022-05-21 & 59720 & 1 & A & 1.3 & 3-bit/8-bit-v1 & J0552+0313, 3C147, 3C147 & 1.272 & No\\
22A-268 & 2022-06-24 & 59754 & 1 & A & 1.4 & 3-bit/8-bit-v1 & J0552+0313, 3C147, 3C147 & 1.041 & No\\
22A-268 & 2022-07-23 & 59783 & 1 & D & 1.3 & 3-bit/8-bit-v1 & J0552+0313, 3C147, 3C147 & 1.028 & No\\
22A-268 & 2022-08-20 & 59811 & 1 & D & 1.3 & 3-bit/8-bit-v1 & J0552+0313, 3C147, 3C147 & 0.986 & No\\
22A-268 & 2022-09-20 & 59842 & 1 & D & 1.3 & 3-bit/8-bit-v1 & J0552+0313, 3C147, 3C147 & 0.976 & No\\
22B-233 & 2022-09-30 & 59852 & 1 & C & 1.3 & 3-bit/8-bit-v1 & J0552+0313, 3C147, 3C147 & 1.015 & No\\
22B-233 & 2022-11-03 & 59886 & 1 & C & 1.3 & 3-bit/8-bit-v1 & J0552+0313, 3C147, 3C147 & 1.049 & No\\
22B-233 & 2022-12-01 & 59914 & 1 & C & 1.3 & 3-bit/8-bit-v1 & J0552+0313, 3C147, 3C147 & 1.058 & No\\
22B-233 & 2022-12-27 & 59940 & 1 & C & 1.3 & 3-bit/8-bit-v1 & J0552+0313, 3C147, 3C147 & 1.023 & No\\
23A-258 & 2023-01-17 & 59961 & 1 & B & 1.3 & 3-bit & J0552+0313, 3C147, 3C147 & 1.018 & Yes\\
23A-258 & 2023-03-21 & 60024 & 1 & B & 1.3 & 3-bit/8-bit-v1 & J0552+0313, 3C147, 3C147 & 1.085 & Yes\\
23A-258 & 2023-03-29 & 60032 & 1 & B & 1.3 & 3-bit/8-bit-v1 & J0552+0313, 3C147, 3C147 & 1.154 & Yes\\
23A-258 & 2023-04-02 & 60036 & 1 & B & 1.3 & 3-bit/8-bit-v1 & J0552+0313, 3C147, 3C147 & 1.087 & Yes\\
23A-258 & 2023-04-30 & 60064 & 1 & B & 1.3 & 3-bit/8-bit-v1 & J0552+0313, 3C147, 3C147 & 0.958 & Yes\\
23A-258 & 2023-05-04 & 60068 & 1 & B & 1.3 & 3-bit/8-bit-v1 & J0552+0313, 3C147, 3C147 & 1.101 & Yes\\
23A-258 & 2023-05-16 & 60080 & 1 & B & 1.3 & 3-bit/8-bit-v1 & J0552+0313, 3C147, 3C147 & 1.073 & Yes\\
23A-258 & 2023-05-19 & 60083 & 1 & B & 1.3 & 3-bit/8-bit-v2 & J0552+0313, 3C147, 3C147 & 0.977 & Yes\\
23A-258 & 2023-08-07 & 60163 & 1 & A & 1.3 & 3-bit/8-bit-v2 & J0552+0313, 3C147, 3C147 & 1.050 & Yes\\
23A-258 & 2023-08-11 & 60167 & 2 & A & 1.3 & 3-bit/8-bit-v2 & J0552+0313, 3C147, 3C147 & 1.014 & Yes\\
23A-258 & 2023-08-19 & 60175 & 1 & A & 1.3 & 3-bit/8-bit-v2 & J0552+0313, 3C147, 3C147 & 1.097 & Yes\\
23A-258 & 2023-08-24 & 60180 & 1 & A & 1.3 & 3-bit/8-bit-v2 & J0552+0313, 3C147, 3C147 & 1.190 & Yes\\
23A-258 & 2023-08-27 & 60183 & 1 & A & 1.3 & 3-bit/8-bit-v2 & J0552+0313, 3C147, 3C147 & 1.091 & Yes\\
23A-258 & 2023-09-17 & 60204 & 2 & A & 1.3 & 3-bit/8-bit-v2 & J0552+0313, 3C147, 3C147 & 1.095 & Yes\\
26A-268 & 2026-02-22 & 61092 & 1 & BnA$\rightarrow$A\tablenotemark{b} & 2.7 & 3-bit/8-bit-v2 & J0552+0313, 3C147, 3C147 & \nodata & Yes\\
\enddata
\tablecomments{The scale factors are calculated using sources within images created using all the EBs from a single configuration and those within a single EB image. The flux ratio of cross-matched sources
is calculated and a Gaussian is fit to the histogram of flux ratios to determine the scale factors. Aside from two outliers, all observations in a given configuration had distributions of flux densities that were within 10\% of each other. The 3-bit sampler setup used basebands centered at 5 GHz and 7 GHz, the 3-bit/8-bit-v1 used 3-bit basebands centered at 5 GHz and 7 GHz and an 8-bit baseband centered at 7.4 GHz, and 3-bit/8-bit-v2 used 3-bit basebands centered at 5 GHz and 7.6 GHz and an 8-bit baseband centered at 6.2 GHz. The standard 3-bit setup used in the B-configuration observation on 2023-01-17 was due to and oversight in the observational setup.}
\tablenotetext{a}{There were 21 antennas moved to their D-configuration pads and the remaining antennas on A-configuration pads were not included in imaging.}
\tablenotetext{b}{Only two antennas were not in their A-configuration positions at the time of observation, having a negligible impact on the resolution.}
\label{vla_cband_obs}

\end{deluxetable}

\begin{deluxetable}{llllcclll}
\label{vla_kband_obs}

\tabletypesize{\scriptsize}
\tablewidth{0pt}
\tablecaption{VLA K-band Observation Log}

\tablehead{\colhead{Proposal Code} & \colhead{Date} & \colhead{Date} & \colhead{EBs} & \colhead{Config.} & \colhead{Duration} & \colhead{Samplers} & \colhead{Calibrators}  & \colhead{Selfcal?}\\
 & \colhead{(YYYY-MM-DD)} &  \colhead{(MJD)} &        &      & \colhead{(hr)}   &  &\colhead{(Complex Gain, Bandpass, Flux)}  &
}
\startdata
15B-229 & 2015-10-17 & 57312 & 1 & D & 1.6 & 8-bit & J0541-0541, 3C84, 3C147 & No\\
15B-229 & 2015-11-21 & 57347 & 1 & D & 1.6 & 8-bit & J0541-0541, 3C84, 3C147 & No\\
21A-409 & 2021-04-05 & 59309 & 1 & D & 1.2 & 3-bit & J0552+0313, 3C84, 3C147 & No\\
22A-268 & 2022-02-18 & 59628 & 1 & A & 1.1 & 8-bit & J0552+0313, 3C84, 3C147 & No\\
22A-268 & 2022-03-18 & 59656 & 1 & A & 1.1 & 8-bit & J0552+0313, 3C84, 3C147 & No\\
22A-268 & 2022-04-02 & 59671 & 1 & A & 1.1 & 8-bit & J0552+0313, 3C84, 3C147 & No\\
22A-268 & 2022-05-13 & 59712 & 1 & A & 1.1 & 8-bit & J0552+0313, 3C84, 3C147 & No\\
22A-268 & 2022-07-25 & 59785 & 1 & D & 1.1 & 8-bit & J0552+0313, 3C84, 3C147 & No\\
22A-268 & 2022-08-22 & 59813 & 1 & D & 1.1 & 8-bit & J0552+0313, 3C84, 3C147 & No\\
22A-268 & 2022-09-20 & 59842 & 1 & D & 1.1 & 8-bit & J0552+0313, 3C84, 3C147 & No\\
22B-233 & 2022-10-02 & 59854 & 1 & C & 1.1 & 8-bit & J0552+0313, 3C84, 3C147 & No\\
22B-233 & 2022-11-17 & 59900 & 1 & C & 1.1 & 8-bit & J0552+0313, 3C84, 3C147 & No\\
22B-233 & 2022-12-01 & 59914 & 1 & C & 1.1 & 8-bit & J0552+0313, 3C84, 3C147 & No\\
22B-233 & 2022-12-27 & 59940 & 1 & C & 1.1 & 8-bit & J0552+0313, 3C84, 3C147 & No\\
23A-258 & 2023-02-07 & 59982 & 1 & B & 1.1 & 8-bit & J0552+0313, 3C84, 3C147 & No\\
23A-258 & 2023-03-05 & 60008 & 1 & B & 1.1 & 8-bit & J0552+0313, 3C84, 3C147 & No\\
23A-258 & 2023-04-06 & 60040 & 1 & B & 1.1 & 8-bit & J0552+0313, 3C84, 3C147 & No\\
23A-258 & 2023-05-20 & 60084 & 1 & B & 1.1 & 8-bit & J0552+0313, 3C84, 3C147 & No\\
23A-258 & 2023-05-29 & 60093 & 1 & B & 1.1 & 8-bit & J0552+0313, 3C84, 3C147 & No\\
23A-258 & 2023-07-24 & 60149 & 1 & A & 1.1 & 8-bit & J0552+0313, 3C84, 3C147 & No\\
23A-258 & 2023-08-19 & 60175 & 1 & A & 1.1 & 8-bit & J0552+0313, 3C84, 3C147 & No\\
23A-258 & 2023-09-09 & 60196 & 1 & A & 1.1 & 8-bit & J0552+0313, 3C84, 3C147 & No\\
26A-268 & 2026-02-02 & 61073 & 1 & A & 1.1 & 8-bit & J0552+0313, 3C84, 3C147 & No\\
\enddata

\end{deluxetable}

\begin{deluxetable}{llll}
\label{relative-scales}
\tabletypesize{\scriptsize}
\tablewidth{0pt}
\tablecaption{Possible Systematic Offsets in the 5~cm Flux Density Scale}

\tablehead{\colhead{Data Set} & \colhead{Reference Data Set} & \colhead{Scale Factor} & \colhead{Cumulative Scale Factor}\tablenotemark{a}\\
}
\startdata
D configuration 2015 & D configuration 2021 &  0.850 & \nodata\\
D configuration 2021 & D configuration 2021 &  1.0 & \nodata \\
C configuration 2021 & D configuration 2021 & 0.984 & \nodata \\ 
B configuration 2021/2022 & C configuration 2021 & 0.941 & 0.926 \\
A configuration 2022 & B configuration 2021/2022 & 0.902 & 0.835\\
D configuration 2022 & D configuration 2021 & 0.960 & \nodata \\
C configuration 2022 & C configuration 2021 & 1.081 & \nodata \\
B-configuration 2023 & B-configuration 2021/2022 & 1.061 & \nodata \\
A-configuration 2023 & A-configuration 2022 & 0.955 & \nodata \\
A-configuration 2026  & A-configuration 2022 & 1.03 & \nodata \\
\enddata
\tablecomments{We compared the relative flux densities of sources identified in an image 
generated using all data taken while the VLA was in a given configuration. The flux ratio of cross-matched sources is calculated and a Gaussian is fit to the histogram of flux ratios to determine the scale factors. Comparisons across configurations were performed using images from the higher resolution configuration tapered and then restored to an identical beam as the lower resolution configuration. Because A-configuration data cannot make an image at D-configuration resolution with reasonable sensitivity, the data from 2022 and 2023, where each configuration was revisited, have their flux densities compared to the same configuration of earlier observations. The observations generally agree within 10\%. The trend of having scale factors less than 1 for more extended configurations may be related to the spatial filtering of flux because decorrelation is generally not significant in the observations.
}
\tablenotetext{a}{This scale factor relates the 2021/2022 A and B configuration observations back to D configuration in 2021 by multiplying all the scale factors together. }

\end{deluxetable}

\begin{deluxetable}{llllllll}
\label{per-epoch-imfit-fluxes}
\tabletypesize{\tiny}
\tablewidth{0pt}
\tablecaption{Per-Epoch Flux Densities}

\tablehead{\colhead{Modified Julian Date} & \colhead{Config.} & \colhead{Type} & \colhead{HOPS-373} & \colhead{HOPS-373-SW} & \colhead{HOPS-373-NE} & \colhead{HOPS-403} & \colhead{HOPS-321} \\
\colhead{(MJD)} & \colhead{}  & \colhead{} & \colhead{(mJy)} & \colhead{(mJy)}  &\colhead{(mJy)} & \colhead{(mJy)} & \colhead{(mJy)} \\
}
\startdata
57301.0 & D & single & 0.032$\pm$0.008 & $<$0.032 & $<$0.032 & 0.058$\pm$0.008 & 0.082$\pm$0.008\\
59297.0 & D & single & 0.047$\pm$0.006 & $<$0.047 & $<$0.047 & 0.044$\pm$0.006 & 0.100$\pm$0.006\\
59324.0 & D & average & 0.045$\pm$0.005 & $<$0.045 & $<$0.045 & 0.056$\pm$0.005 & 0.084$\pm$0.005\\
59349.0 & D & single & 0.045$\pm$0.006 & $<$0.045 & $<$0.045 & 0.058$\pm$0.006 & 0.070$\pm$0.006\\
59420.0 & C & single & 0.036$\pm$0.006 & $<$0.036 & $<$0.036 & 0.046$\pm$0.006 & 0.069$\pm$0.006\\
59434.0 & C & single & 0.036$\pm$0.006 & $<$0.036 & $<$0.036 & 0.051$\pm$0.006 & 0.073$\pm$0.006\\
59450.0 & C & single & 0.032$\pm$0.005 & $<$0.034 & $<$0.032 & 0.050$\pm$0.005 & 0.075$\pm$0.005\\
59450.0 & C & average & 0.034$\pm$0.003 & $<$0.032 & $<$0.034 & 0.046$\pm$0.003 & 0.075$\pm$0.003\\
59468.0 & C & single & 0.039$\pm$0.005 & $<$0.039 & $<$0.039 & 0.039$\pm$0.005 & 0.090$\pm$0.005\\
59502.0 & B & single & 0.034$\pm$0.006 & 0.018$\pm$0.004 & 0.016$\pm$0.004 & 0.044$\pm$0.004 & 0.079$\pm$0.004\\
59524.0 & B & single & 0.032$\pm$0.006 & 0.018$\pm$0.004 & 0.014$\pm$0.004 & 0.044$\pm$0.004 & 0.074$\pm$0.004\\
59548.0 & B & average & 0.029$\pm$0.003 & 0.020$\pm$0.002 & 0.009$\pm$0.002 & 0.044$\pm$0.002 & 0.078$\pm$0.002\\
59555.0 & B & single & 0.032$\pm$0.006 & 0.025$\pm$0.004 & $<$0.012 & 0.042$\pm$0.004 & 0.071$\pm$0.004\\
59578.0 & B & single & 0.045$\pm$0.006 & 0.027$\pm$0.004 & $<$0.011 & 0.038$\pm$0.004 & 0.078$\pm$0.004\\
59591.0 & B & single & 0.027$\pm$0.004 & 0.019$\pm$0.003 & 0.008$\pm$0.003 & 0.048$\pm$0.003 & 0.088$\pm$0.003\\
59643.0 & A & single & 0.034$\pm$0.006 & 0.019$\pm$0.004 & 0.015$\pm$0.004 & 0.044$\pm$0.004 & 0.048$\pm$0.004\\
59676.0 & A & single & 0.024$\pm$0.006 & 0.015$\pm$0.004 & 0.009$\pm$0.004 & 0.047$\pm$0.004 & 0.060$\pm$0.004\\
59700.0 & A & average & 0.024$\pm$0.003 & 0.014$\pm$0.002 & 0.010$\pm$0.002 & 0.041$\pm$0.002 & 0.064$\pm$0.002\\
59720.0 & A & single & 0.030$\pm$0.006 & 0.020$\pm$0.004 & $<$0.011 & 0.046$\pm$0.004 & 0.080$\pm$0.004\\
59754.0 & A & single & 0.030$\pm$0.006 & 0.012$\pm$0.004 & 0.018$\pm$0.004 & 0.043$\pm$0.004 & 0.069$\pm$0.004\\
59783.0 & D & single & 0.047$\pm$0.006 & $<$0.047 & $<$0.047 & 0.041$\pm$0.006 & 0.094$\pm$0.006\\
59811.0 & D & single & 0.033$\pm$0.007 & $<$0.033 & $<$0.043 & 0.049$\pm$0.007 & 0.094$\pm$0.007\\
59811.0 & D & average & 0.043$\pm$0.005 & $<$0.043 & $<$0.033 & 0.049$\pm$0.005 & 0.091$\pm$0.005\\
59842.0 & D & single & 0.039$\pm$0.006 & $<$0.039 & $<$0.039 & 0.041$\pm$0.006 & 0.089$\pm$0.006\\
59852.0 & C & single & 0.030$\pm$0.005 & $<$0.030 & $<$0.030 & 0.052$\pm$0.005 & 0.098$\pm$0.005\\
59886.0 & C & single & 0.031$\pm$0.005 & $<$0.031 & $<$0.031 & 0.054$\pm$0.005 & 0.121$\pm$0.005\\
59898.0 & C & average & 0.030$\pm$0.003 & $<$0.030 & $<$0.030 & 0.047$\pm$0.003 & 0.123$\pm$0.003\\
59914.0 & C & single & 0.031$\pm$0.005 & $<$0.031 & $<$0.031 & 0.032$\pm$0.005 & 0.133$\pm$0.005\\
59940.0 & C & single & 0.031$\pm$0.004 & $<$0.031 & $<$0.031 & 0.042$\pm$0.004 & 0.138$\pm$0.004\\
59961.0 & B & single & 0.027$\pm$0.004 & 0.014$\pm$0.003 & $<$0.009 & 0.036$\pm$0.003 & 0.126$\pm$0.003\\
60024.0 & B & single & 0.036$\pm$0.006 & 0.026$\pm$0.004 & 0.010$\pm$0.004 & 0.050$\pm$0.004 & 0.121$\pm$0.004\\
60032.0 & B & single & 0.038$\pm$0.006 & 0.023$\pm$0.004 & $<$0.012 & 0.050$\pm$0.004 & 0.123$\pm$0.004\\
60033.0 & B & average & 0.031$\pm$0.001 & 0.018$\pm$0.001 & 0.013$\pm$0.001 & 0.042$\pm$0.001 & 0.118$\pm$0.001\\
60036.0 & B & single & 0.033$\pm$0.004 & 0.021$\pm$0.003 & 0.012$\pm$0.003 & 0.040$\pm$0.003 & 0.136$\pm$0.003\\
60064.0 & B & single & 0.031$\pm$0.006 & 0.016$\pm$0.004 & 0.015$\pm$0.004 & 0.042$\pm$0.004 & 0.105$\pm$0.004\\
60068.0 & B & single & 0.043$\pm$0.006 & 0.021$\pm$0.004 & $<$0.013 & 0.042$\pm$0.004 & 0.105$\pm$0.004\\
60080.0 & B & single & 0.021$\pm$0.004 & 0.021$\pm$0.004 & $<$0.012 & 0.049$\pm$0.004 & 0.108$\pm$0.004\\
60083.0 & B & single & 0.018$\pm$0.007 & 0.011$\pm$0.005 & $<$0.014 & 0.043$\pm$0.005 & 0.094$\pm$0.005\\
60163.0 & A & single & 0.016$\pm$0.004 & 0.016$\pm$0.004 & $<$0.011 & 0.045$\pm$0.004 & 0.126$\pm$0.004\\
60167.0 & A & single & 0.014$\pm$0.003 & 0.014$\pm$0.003 & $<$0.007 & 0.035$\pm$0.003 & 0.140$\pm$0.003\\
60175.0 & A & single & 0.024$\pm$0.004 & 0.024$\pm$0.004 & $<$0.011 & 0.051$\pm$0.004 & 0.145$\pm$0.004\\
60180.0 & A & average & 0.019$\pm$0.001 & 0.024$\pm$0.004 & 0.013$\pm$0.004 & 0.034$\pm$0.001 & 0.129$\pm$0.001\\
60180.0 & A & single & 0.037$\pm$0.006 & 0.015$\pm$0.001 & 0.004$\pm$0.001 & 0.049$\pm$0.004 & 0.153$\pm$0.004\\
60183.0 & A & single & 0.027$\pm$0.004 & 0.017$\pm$0.003 & 0.010$\pm$0.003 & 0.033$\pm$0.003 & 0.140$\pm$0.003\\
60204.0 & A & single & 0.019$\pm$0.003 & 0.019$\pm$0.003 & $<$0.010 & 0.038$\pm$0.003 & 0.134$\pm$0.003\\
61093.0 & A & single & 0.037$\pm$0.003 & 0.022$\pm$0.002 & 0.015$\pm$0.002 & 0.044$\pm$0.002 & 0.143$\pm$0.002\\

\enddata
\tablecomments{Measurements with type 'single' refers to a single observation while 'average' refers to the flux densities measured from an image that uses all data within a particular configuration. HOPS-373-SW and -NE are unresolved from each other at D and C configuration, therefore they are listed as upper limits in those configurations.}
\end{deluxetable}

\begin{deluxetable}{lllll}
\label{k_band_fluxes}
\tabletypesize{\scriptsize}
\tablewidth{0pt}
\tablecaption{1.3~cm Flux Density Measurements}

\tablehead{\colhead{Date} & \colhead{Date} & \colhead{Configuration} & \colhead{HOPS-373 SW} & \colhead{HOPS-373 NE}\\
\colhead{(UT)} & \colhead{(MJD)} & \colhead{} & \colhead{($\mu$Jy)} & \colhead{($\mu$Jy)}
}
\startdata
2015-10-17 & 57312 & D & 146$\pm$9.4 & 170$\pm$9.4\\
2015-11-21 & 57347 & D & 160$\pm$13.4 & 169$\pm$13.4\\
Average & 57330 & D & 152$\pm$9.6 & 170$\pm$9.6\\
2021-04-05 & 59309 & D & 151$\pm$11.5 & 145$\pm$11.5\\
2022-02-18 & 59628 & BnA & 94$\pm$13.2 & 110$\pm$13.2\\
2022-03-18 & 59656 & A & 85$\pm$14.6 & 103$\pm$14.6\\
2022-04-02 & 59671 & A & 67$\pm$12.7 & 67$\pm$12.7\\
2022-05-13 & 59712 & A & 94$\pm$14.0 & 65$\pm$14.0\\
Average & 59667 & A & 84$\pm$6.8 & 87$\pm$6.8\\
2022-07-25 & 59785 & D & 226$\pm$32.8 & 227$\pm$32.8\\
2022-08-22 & 59813 & D & 146$\pm$30.4 & 161$\pm$30.4\\
2022-09-20 & 59842 & D & 181$\pm$35.9 & 164$\pm$35.9\\
Average & 59814 & D & 177$\pm$21.4 & 182$\pm$21.4\\
2022-10-02 & 59854 & C & 161$\pm$25.4 & 151$\pm$25.4\\
2022-11-17 & 59900 & C & 126$\pm$16.1 & 156$\pm$16.1\\
2022-12-01 & 59914 & C & 138$\pm$19.6 & 162$\pm$19.6\\
2022-12-27 & 59940 & C & 151$\pm$19.4 & 155$\pm$19.4\\
Average & 59902 & C & 134$\pm$14.7 & 154$\pm$14.7\\
2023-02-07 & 59982 & B & 93$\pm$8.9 & 121$\pm$8.9\\
2023-03-05 & 60008 & B & 94$\pm$10.8 & 134$\pm$10.8\\
2023-04-06 & 60040 & B & 126$\pm$13.5 & 125$\pm$13.5\\
2023-05-20 & 60084 & B & 71$\pm$22.5 & 127$\pm$22.5\\
2023-05-29 & 60093 & B & 116$\pm$14.0 & 84$\pm$14.0\\
Average & 60041 & B & 101$\pm$6.7 & 118$\pm$6.7\\
2023-07-24 & 60149 & A & 68$\pm$16.6 & 84$\pm$16.6\\
2023-08-19 & 60175 & A & 94$\pm$17.6 & 60$\pm$17.6\\
2023-09-09 & 60196 & A & 123$\pm$20.2 & 105$\pm$20.2\\
2026-02-02 & 60174 & BnA & 93$\pm$12.7 & 78$\pm$12.7\\
\enddata

\end{deluxetable}

\clearpage
\begin{deluxetable}{llllllllc}
\label{maser-properties}
\tabletypesize{\scriptsize}
\tablewidth{0pt}
\tablecaption{Water Maser Properties}

\tablehead{\colhead{Date} & \colhead{Date} & \colhead{Config.} & \colhead{RA} & \colhead{Dec}  & \colhead{Flux Density} & \colhead{V$_{lsr}$} & \colhead{FWHM} & \colhead{RMS}\\
\colhead{(YYYY-MM-DD)}    & \colhead{(MJD)}    &  & \colhead{(ICRS)} & \colhead{(ICRS)}             &  \colhead{(Jy)}       & \colhead{(km s$^{-1}$)}     & \colhead{(km s$^{-1}$)} & \colhead{(Jy~beam$^{-1}$)} 
}
\startdata
2015-10-17 & 57312 & D & 05:46:30.9217$\pm$0.333 & -00:02:35.504$\pm$0.202 & 0.005$\pm$0.006 & -15.2$\pm$0.17 & 0.59$\pm$0.51 & 0.0030\\
2015-11-21 & 57347 & D & 05:46:30.8895$\pm$1.076 & -00:02:33.4687$\pm$0.129 & 0.009$\pm$0.012 & -14.8$\pm$0.14 & 0.59$\pm$0.58 & 0.0059\\
2021-04-05 & 59309 & D & 05:46:30.8701$\pm$0.202 & -00:02:35.0366$\pm$0.064 & 0.053$\pm$0.018 & 9.8$\pm$0.07 & 0.82$\pm$0.23 & 0.0122\\
2022-02-18 & 59628 & A & 05:46:30.9004$\pm$0.001 & -00:02:35.1696$\pm$0.002 & 0.221$\pm$0.009 & 3.3$\pm$0.01 & 0.96$\pm$0.03 & 0.0041\\
2022-03-18 & 59656 & A & 05:46:30.9013$\pm$0.002 & -00:02:35.1649$\pm$0.002 & 0.080$\pm$0.007 & 3.3$\pm$0.03 & 1.17$\pm$0.07 & 0.0048\\
2022-04-02 & 59671 & A & 05:46:30.9009$\pm$0.006 & -00:02:35.1573$\pm$0.004 & 0.036$\pm$0.007 & 3.0$\pm$0.06 & 1.02$\pm$0.14 & 0.0063\\
2022-05-13 & 59712 & A & 05:46:30.9021$\pm$0.009 & -00:02:35.153$\pm$0.010 & 0.041$\pm$0.008 & 3.0$\pm$0.09 & 1.55$\pm$0.22 & 0.0057\\
2022-07-25 & 59785 & D & 05:46:30.5768$\pm$0.789 & -00:02:36.4675$\pm$0.506 & 0.020$\pm$0.005 & 3.5$\pm$0.08 & 0.86$\pm$0.17 & 0.0081\\
2022-08-22 & 59813 & D & 05:46:30.9714$\pm$0.299 & -00:02:37.2819$\pm$0.176 & 0.031$\pm$0.010 & -15.3$\pm$0.16 & 1.60$\pm$0.39 & 0.0086\\
2022-09-20 & 59842 & D & \nodata & \nodata & \nodata & \nodata & \nodata & 0.0081\\
2022-10-02 & 59854 & C & \nodata & \nodata & \nodata & \nodata & \nodata & 0.0066\\
2022-11-17 & 59900 & C & \nodata & \nodata & \nodata & \nodata & \nodata & 0.0022\\
2022-12-01 & 59914 & C & \nodata & \nodata & \nodata & \nodata & \nodata & 0.0037\\
2022-12-27 & 59940 & C & \nodata & \nodata & \nodata & \nodata & \nodata & 0.0039\\
2023-02-07 & 59982 & B & \nodata & \nodata & \nodata & \nodata & \nodata & 0.0053\\
2023-03-05 & 60008 & B & \nodata & \nodata & \nodata & \nodata & \nodata & 0.0047\\
2023-04-06 & 60040 & B & \nodata & \nodata & \nodata & \nodata & \nodata & 0.0061\\
2023-05-20 & 60084 & B & 05:46:30.9026$\pm$0.017 & -00:02:35.2349$\pm$0.011 & 0.100$\pm$0.006 & 20.6$\pm$0.02 & 0.93$\pm$0.04 & 0.0096\\
2023-05-29 & 60093 & B & 05:46:30.8984$\pm$0.263 & -00:02:35.2082$\pm$0.263 & 0.016$\pm$0.031 & -10.0$\pm$0.11 & 1.12$\pm$0.27 & 0.0071\\
2023-05-29 & 60093 & B & 05:46:30.8988$\pm$0.067 & -00:02:35.1162$\pm$0.067  & 0.019$\pm$0.027 & -8.0$\pm$0.10 & 1.17$\pm$0.25 & 0.0071\\
2023-05-29 & 60093 & B & 05:46:30.9061$\pm$0.008 & -00:02:35.2005$\pm$0.006 & 0.052$\pm$0.009 & 20.6$\pm$0.06 & 1.00$\pm$0.13 & 0.0061\\
2023-07-24 & 60149 & A & 05:46:30.8985$\pm$0.003 & -00:02:35.1518$\pm$0.003 & 0.049$\pm$0.010 & -10.3$\pm$0.07 & 1.12$\pm$0.18 & 0.0090\\
2023-07-24 & 60149 & A & 05:46:30.8976$\pm$0.008 & -00:02:35.1567$\pm$0.008  & 0.022$\pm$0.021 & -8.7$\pm$0.10 & 0.88$\pm$0.29 & 0.0090\\
2023-07-24 & 60149 & A & 05:46:30.9042$\pm$0.024 & -00:02:35.1841$\pm$0.022 & 0.023$\pm$0.012 & 20.1$\pm$0.27 & 1.56$\pm$0.63 & 0.0088\\
2023-08-19 & 60175 & A & 05:46:30.899$\pm$0.005 & -00:02:35.1397$\pm$0.004 & 0.075$\pm$0.011 & -10.4$\pm$0.07 & 1.50$\pm$0.16 & 0.0071\\
2023-08-19 & 60175 & A & 05:46:30.9048$\pm$0.029 & -00:02:35.2103$\pm$0.024 & 0.016$\pm$0.010 & 20.9$\pm$0.19 & 1.01$\pm$0.44 & 0.0074\\
2023-09-09 & 60196 & A & 05:46:30.8988$\pm$0.005 & -00:02:35.1465$\pm$0.005 & 0.077$\pm$0.012 & -10.4$\pm$0.05 & 1.13$\pm$0.13 & 0.0074\\
2023-09-09 & 60196 & A & 05:46:30.9054$\pm$0.003 & -00:02:35.1893$\pm$0.002 & 0.050$\pm$0.008 & 21.3$\pm$0.06 & 1.08$\pm$0.14 & 0.0077\\
\enddata
\tablecomments{Fitted properties of water maser spots observed toward HOPS-373. The flux density, V$_{lsr}$ and FWHM are from Gaussian fitting to the spectrum, while their positions are derived from image-plane Gaussian fits to peak intensity maps generated over the velocity range of the maser.}
\end{deluxetable}

\begin{deluxetable}{lllll}
\label{radio-spectrum-data}
\tabletypesize{\scriptsize}
\tablewidth{0pt}
\tablecaption{Continuum Flux Density Measurements}
\label{radio-spectrum-fluxes}
\tablehead{\colhead{Wavelength} & \colhead{Resolution} & \colhead{HOPS-373 NE} & \colhead{HOPS-373 SW} & \colhead{References}\\
\colhead{(mm)} & \colhead{(\arcsec)} & \colhead{(mJy)} & \colhead{(mJy)} &
}
\startdata
0.87 & 4.99$\times$2.90 & 325.000$\pm$40& 406$\pm$46 & 1, 2\\
0.87 & 0.13$\times$0.10 & 103.8$\pm$1.6& 92.21$\pm$1.0 & 3\\
0.87 & 0.08$\times$0.07 & 75.1$\pm$2.3 & 90.4$\pm$1.4 & 4\\
2.8 & 2.29$\times$2.04 & 25.8$\pm$0.11& 23.2$\pm$0.15 & 1, 5\\
8.1 & 2.24$\times$1.55 & 0.780$\pm$0.031& 0.511$\pm$0.031 & 1\\
9.0 & 3.48$\times$2.41 & 0.569$\pm$0.021& 0.391$\pm$0.021 & 1\\
10.0 & 3.73$\times$2.51 & 0.385$\pm$0.018& 0.299$\pm$0.018 & 1\\
13.0 & 3.04$\times$2.42 & 0.175$\pm$0.027& 0.162$\pm$0.027 & 1\\
42.8 & 1.35$\times$0.99 & 0.011$\pm$0.001& 0.020$\pm$0.001 & 1\\
50.0 & 1.50$\times$1.11 & 0.010$\pm$0.001& 0.018$\pm$0.001 & 1\\
60.0 & 1.85$\times$1.40 & 0.008$\pm$0.002& 0.015$\pm$0.002 & 1\\
\enddata
\tablecomments{References: 1. This work, 2. \citet{federman2023}, 3. \citet{tobin2020}, 4. \citet{lee2023}, and 5. \citet{tobin2015}. When both this work and another reference is provided, the flux density was measured as part of this work, but the data were originally presented in the referenced work.}

\end{deluxetable}

\clearpage
\startlongtable
\begin{deluxetable}{lllllll}
\label{c-band-catalog}
\tabletypesize{\tiny}
\tablewidth{0pt}
\tablecaption{Primary Source Catalog}
\tablehead{\colhead{Source} & \colhead{RA} & \colhead{DEC}   & \colhead{Flux Density} & \colhead{Peak Intensity}  & \colhead{Blended Sources\tablenotemark{a}} & \colhead{Other Names}\\
\colhead{}    & \colhead{(ICRS)} & \colhead{(ICRS)}         &  \colhead{($\mu$Jy)}       & \colhead{($\mu$Jy~bm$^{-1}$)}  &  & 
}
\startdata
NGC-2068-B-VLA-1 & 05:46:31.846$\pm$0.15 & +00:02:03.08$\pm$0.23 & 44.0$\pm$14.1 & 29.1$\pm$6.1 & -- & None\\
NGC-2068-B-VLA-2 & 05:46:28.188$\pm$0.13 & +00:01:57.58$\pm$0.14 & 62.7$\pm$13.2 & 38.9$\pm$5.4 & -- & None\\
NGC-2068-B-VLA-3 & 05:46:44.142$\pm$0.02 & +00:01:40.81$\pm$0.02 & 921.2$\pm$31.2 & 626.9$\pm$13.7 & -- & NVSS J054644+000144\\
NGC-2068-B-VLA-4 & 05:46:24.798$\pm$0.36 & +00:01:20.27$\pm$0.26 & 62.2$\pm$17.7 & 18.1$\pm$4.1 & -- & None\\
NGC-2068-B-VLA-5 & 05:46:31.268$\pm$0.05 & +00:01:10.46$\pm$0.08 & 47.7$\pm$6.4 & 44.2$\pm$3.5 & -- & None\\
NGC-2068-B-VLA-6 & 05:46:47.555$\pm$0.04 & +00:00:54.30$\pm$0.06 & 202.3$\pm$22.4 & 179.3$\pm$11.9 & -- & None\\
NGC-2068-B-VLA-7 & 05:46:34.586$\pm$0.03 & +00:00:52.45$\pm$0.05 & 75.3$\pm$5.7 & 63.0$\pm$2.9 & -- & None\\
NGC-2068-B-VLA-8 & 05:46:43.118$\pm$0.03 & +00:00:52.44$\pm$0.04 & 236.4$\pm$13.9 & 183.9$\pm$6.7 & 9-D & HOPS-363; 2MASS J05464311+0000523; SSV LDN 1630 19\\
NGC-2068-B-VLA-9 & 05:46:43.770$\pm$0.11 & +00:00:51.14$\pm$0.25 & 148.2$\pm$23.8 & 51.6$\pm$6.3 & 8-D & None\\
NGC-2068-B-VLA-10 & 05:46:38.719$\pm$0.08 & +00:00:34.59$\pm$0.18 & 21.5$\pm$5.8 & 19.0$\pm$3.0 & -- & None\\
NGC-2068-B-VLA-11 & 05:46:37.560$\pm$0.07 & +00:00:33.84$\pm$0.10 & 45.4$\pm$6.8 & 33.1$\pm$3.2 & -- & HOPS-324; 2MASS J05463755+0000338\\
NGC-2068-B-VLA-12 & 05:46:47.014$\pm$0.14 & +00:00:27.35$\pm$0.20 & 64.3$\pm$19.6 & 46.6$\pm$9.0 & 13-D & 2MASS J05464703+0000265\\
NGC-2068-B-VLA-13 & 05:46:47.672$\pm$0.07 & +00:00:25.12$\pm$0.12 & 102.1$\pm$17.8 & 76.8$\pm$8.4 & 12-D & HGBS J054647.3+000026; 2MASS J05464775+0000242; 2MASS J05464776+0000243\\
NGC-2068-B-VLA-14 & 05:46:33.168$\pm$0.01 & +00:00:02.09$\pm$0.02 & 125.6$\pm$4.1 & 111.2$\pm$2.2 & -- & HOPS-321; 2MASS J05463318+0000021; [G99] LBS 17 cm1\\
NGC-2068-B-VLA-15 & 05:46:16.093$\pm$0.03 & -00:00:09.11$\pm$0.06 & 138.1$\pm$11.3 & 106.1$\pm$5.4 & -- & None\\
NGC-2068-B-VLA-16 & 05:46:18.585$\pm$0.12 & -00:00:19.31$\pm$0.13 & 127.4$\pm$15.0 & 39.2$\pm$3.6 & -- & 2MASS J05461858-0000190\\
NGC-2068-B-VLA-17 & 05:46:32.727$\pm$0.14 & -00:00:21.20$\pm$0.18 & 19.7$\pm$4.8 & 12.7$\pm$2.0 & -- & [MAW2001] LBS17-MM11\\
NGC-2068-B-VLA-18 & 05:46:51.643$\pm$0.09 & -00:00:28.15$\pm$0.15 & 82.0$\pm$20.6 & 72.3$\pm$10.8 & -- & None\\
NGC-2068-B-VLA-19 & 05:46:44.732$\pm$0.02 & -00:00:28.38$\pm$0.04 & 110.0$\pm$7.2 & 104.1$\pm$4.0 & -- & None\\
NGC-2068-B-VLA-20 & 05:46:37.769$\pm$0.21 & -00:00:32.10$\pm$0.19 & 16.8$\pm$5.3 & 10.2$\pm$2.2 & -- & None\\
NGC-2068-B-VLA-21 & 05:46:24.775$\pm$0.12 & -00:00:38.30$\pm$0.22 & 14.3$\pm$4.2 & 10.6$\pm$2.0 & -- & None\\
NGC-2068-B-VLA-22 & 05:46:29.417$\pm$0.02 & -00:00:50.01$\pm$0.03 & 65.9$\pm$3.2 & 53.7$\pm$1.6 & -- & None\\
NGC-2068-B-VLA-23 & 05:46:27.902$\pm$0.02 & -00:00:52.15$\pm$0.04 & 42.9$\pm$2.8 & 38.9$\pm$1.5 & -- & HOPS-403; HGBS J054627.8-000053\\
NGC-2068-B-VLA-24 & 05:46:18.735$\pm$0.12 & -00:00:57.13$\pm$0.19 & 24.7$\pm$6.8 & 18.6$\pm$3.2 & -- & None\\
NGC-2068-B-VLA-25 & 05:46:46.375$\pm$0.21 & -00:01:00.40$\pm$0.21 & 35.7$\pm$10.8 & 20.1$\pm$4.2 & -- & None\\
NGC-2068-B-VLA-26 & 05:46:37.967$\pm$0.07 & -00:01:02.95$\pm$0.12 & 20.9$\pm$3.9 & 16.9$\pm$1.9 & -- & None\\
NGC-2068-B-VLA-27 & 05:46:33.936$\pm$0.01 & -00:01:07.35$\pm$0.02 & 111.0$\pm$3.0 & 103.1$\pm$1.6 & -- & None\\
NGC-2068-B-VLA-28 & 05:46:32.760$\pm$0.01 & -00:01:07.68$\pm$0.02 & 67.7$\pm$2.8 & 61.3$\pm$1.5 & -- & None\\
NGC-2068-B-VLA-29 & 05:46:40.387$\pm$0.13 & -00:01:15.39$\pm$0.23 & 17.7$\pm$5.0 & 11.2$\pm$2.1 & -- & None\\
NGC-2068-B-VLA-30 & 05:46:24.961$\pm$0.08 & -00:01:15.65$\pm$0.11 & 12.7$\pm$2.9 & 13.2$\pm$1.7 & -- & None\\
NGC-2068-B-VLA-31 & 05:46:16.355$\pm$0.14 & -00:01:18.46$\pm$0.20 & 21.8$\pm$7.2 & 17.8$\pm$3.6 & -- & [SAR2004] 18\\
NGC-2068-B-VLA-32 & 05:46:35.690$\pm$0.05 & -00:01:22.11$\pm$0.09 & 21.5$\pm$3.2 & 18.9$\pm$1.7 & -- & None\\
NGC-2068-B-VLA-33 & 05:46:08.950$\pm$0.10 & -00:01:23.60$\pm$0.13 & 113.6$\pm$26.1 & 94.5$\pm$13.2 & -- & None\\
NGC-2068-B-VLA-34 & 05:46:19.539$\pm$0.03 & -00:01:29.07$\pm$0.05 & 48.7$\pm$4.7 & 45.8$\pm$2.6 & -- & None\\
NGC-2068-B-VLA-35 & 05:46:32.434$\pm$0.07 & -00:01:31.67$\pm$0.15 & 24.4$\pm$3.9 & 14.4$\pm$1.5 & 36-D & None\\
NGC-2068-B-VLA-36 & 05:46:32.931$\pm$0.10 & -00:01:34.34$\pm$0.17 & 8.8$\pm$2.6 & 8.6$\pm$1.5 & 35-D & None\\
NGC-2068-B-VLA-37 & 05:46:35.606$\pm$0.02 & -00:01:37.11$\pm$0.03 & 38.9$\pm$2.5 & 39.2$\pm$1.5 & -- & 2MASS J05463560-0001371\\
NGC-2068-B-VLA-38 & 05:46:28.808$\pm$0.12 & -00:01:42.79$\pm$0.31 & 15.1$\pm$3.9 & 7.0$\pm$1.3 & -- & None\\
NGC-2068-B-VLA-39 & 05:46:19.271$\pm$0.09 & -00:01:50.33$\pm$0.18 & 19.5$\pm$5.0 & 15.5$\pm$2.4 & 40-D & None\\
NGC-2068-B-VLA-40 & 05:46:18.701$\pm$0.01 & -00:01:53.21$\pm$0.03 & 88.7$\pm$4.2 & 90.4$\pm$2.4 & 39-D & None\\
NGC-2068-B-VLA-41 & 05:46:36.940$\pm$0.08 & -00:02:05.32$\pm$0.14 & 12.4$\pm$2.9 & 11.3$\pm$1.6 & -- & None\\
NGC-2068-B-VLA-42 & 05:46:49.508$\pm$0.05 & -00:02:05.64$\pm$0.09 & 78.3$\pm$11.0 & 63.1$\pm$5.4 & -- & None\\
NGC-2068-B-VLA-43 & 05:46:12.239$\pm$0.01 & -00:02:07.45$\pm$0.01 & 785.5$\pm$12.4 & 791.0$\pm$7.1 & -- & GBS-VLA J054612.24-000207.7\\
NGC-2068-B-VLA-44 & 05:46:35.556$\pm$0.15 & -00:02:13.10$\pm$0.19 & 12.8$\pm$3.7 & 8.7$\pm$1.6 & -- & None\\
NGC-2068-B-VLA-45 & 05:46:32.420$\pm$0.07 & -00:02:17.27$\pm$0.12 & 22.6$\pm$3.4 & 14.6$\pm$1.4 & -- & None\\
NGC-2068-B-VLA-46 & 05:46:15.470$\pm$0.12 & -00:02:17.77$\pm$0.15 & 29.9$\pm$7.7 & 23.8$\pm$3.8 & -- & None\\
NGC-2068-B-VLA-47 & 05:46:28.975$\pm$0.11 & -00:02:19.31$\pm$0.16 & 10.3$\pm$2.9 & 8.9$\pm$1.5 & -- & None\\
NGC-2068-B-VLA-48 & 05:46:11.592$\pm$0.01 & -00:02:20.19$\pm$0.02 & 406.6$\pm$12.9 & 377.3$\pm$7.0 & -- & 2MASS J05461157-0002202\\
NGC-2068-B-VLA-49 & 05:46:27.174$\pm$0.08 & -00:02:24.30$\pm$0.12 & 12.6$\pm$2.6 & 11.3$\pm$1.4 & 56-D & None\\
NGC-2068-B-VLA-50 & 05:46:19.774$\pm$0.10 & -00:02:30.66$\pm$0.22 & 14.6$\pm$4.4 & 12.0$\pm$2.2 & 53-D,54-D & None\\
NGC-2068-B-VLA-51 & 05:46:31.118$\pm$0.13 & -00:02:33.23$\pm$0.21 & 17.1$\pm$4.3 & 9.5$\pm$1.6 & 52-D,52-C,55-D & HOPS-373; HOPS 373NE; HH 19-27 H2O Maser\\
NGC-2068-B-VLA-52 & 05:46:30.917$\pm$0.07 & -00:02:35.13$\pm$0.10 & 19.7$\pm$3.1 & 16.6$\pm$1.6 & 51-D,51-C & HOPS-373; HOPS 373SW\\
NGC-2068-B-VLA-53 & 05:46:19.983$\pm$0.08 & -00:02:36.14$\pm$0.12 & 26.9$\pm$5.1 & 20.7$\pm$2.4 & 50-D,54-D & None\\
NGC-2068-B-VLA-54 & 05:46:19.646$\pm$0.15 & -00:02:36.62$\pm$0.17 & 16.1$\pm$5.2 & 13.3$\pm$2.6 & 50-D,53-D & None\\
NGC-2068-B-VLA-55 & 05:46:32.010$\pm$0.10 & -00:02:37.64$\pm$0.25 & 13.6$\pm$3.6 & 8.9$\pm$1.5 & 51-D & None\\
NGC-2068-B-VLA-56 & 05:46:27.134$\pm$0.23 & -00:02:38.01$\pm$0.24 & 14.0$\pm$4.2 & 6.6$\pm$1.4 & 49-D & None\\
NGC-2068-B-VLA-57 & 05:46:46.089$\pm$0.13 & -00:02:42.23$\pm$0.18 & 22.1$\pm$6.5 & 17.6$\pm$3.2 & -- & None\\
NGC-2068-B-VLA-58 & 05:46:34.307$\pm$0.03 & -00:02:50.87$\pm$0.06 & 30.7$\pm$2.9 & 25.7$\pm$1.5 & -- & None\\
NGC-2068-B-VLA-59 & 05:46:30.554$\pm$0.01 & -00:03:00.78$\pm$0.02 & 60.5$\pm$2.3 & 55.6$\pm$1.2 & 60-D & None\\
NGC-2068-B-VLA-60 & 05:46:29.671$\pm$0.17 & -00:03:05.43$\pm$0.15 & 13.0$\pm$3.1 & 7.8$\pm$1.3 & 59-D & None\\
NGC-2068-B-VLA-61 & 05:46:35.035$\pm$0.03 & -00:03:11.50$\pm$0.06 & 33.2$\pm$3.0 & 30.4$\pm$1.6 & -- & None\\
NGC-2068-B-VLA-62 & 05:46:33.804$\pm$0.11 & -00:03:12.53$\pm$0.16 & 15.2$\pm$3.5 & 10.1$\pm$1.5 & 63-D & None\\
NGC-2068-B-VLA-63 & 05:46:33.197$\pm$0.03 & -00:03:13.34$\pm$0.04 & 43.6$\pm$2.9 & 34.6$\pm$1.4 & 62-D & None\\
NGC-2068-B-VLA-64 & 05:46:40.860$\pm$0.09 & -00:03:13.44$\pm$0.14 & 18.7$\pm$4.2 & 15.1$\pm$2.1 & -- & None\\
NGC-2068-B-VLA-65 & 05:46:31.886$\pm$0.05 & -00:03:21.42$\pm$0.11 & 18.0$\pm$2.8 & 14.4$\pm$1.3 & -- & None\\
NGC-2068-B-VLA-66 & 05:46:31.418$\pm$0.07 & -00:03:36.65$\pm$0.12 & 17.1$\pm$3.2 & 13.8$\pm$1.6 & -- & None\\
NGC-2068-B-VLA-67 & 05:46:37.635$\pm$0.13 & -00:03:40.93$\pm$0.20 & 15.8$\pm$4.4 & 10.5$\pm$1.9 & -- & None\\
NGC-2068-B-VLA-68 & 05:46:18.799$\pm$0.10 & -00:03:44.08$\pm$0.16 & 19.5$\pm$5.4 & 17.7$\pm$2.9 & -- & None\\
NGC-2068-B-VLA-69 & 05:46:13.941$\pm$0.11 & -00:03:55.74$\pm$0.22 & 57.8$\pm$13.4 & 32.3$\pm$5.1 & -- & None\\
NGC-2068-B-VLA-70 & 05:46:32.060$\pm$0.06 & -00:04:03.73$\pm$0.10 & 17.9$\pm$3.0 & 15.8$\pm$1.6 & -- & None\\
NGC-2068-B-VLA-71 & 05:46:23.475$\pm$0.01 & -00:04:11.00$\pm$0.02 & 130.1$\pm$3.7 & 120.2$\pm$2.0 & -- & None\\
NGC-2068-B-VLA-72 & 05:46:19.606$\pm$0.10 & -00:04:14.12$\pm$0.16 & 27.6$\pm$6.1 & 19.3$\pm$2.7 & -- & None\\
NGC-2068-B-VLA-73 & 05:46:30.892$\pm$0.08 & -00:04:18.14$\pm$0.12 & 14.9$\pm$3.2 & 13.0$\pm$1.7 & -- & None\\
NGC-2068-B-VLA-74 & 05:46:27.557$\pm$0.09 & -00:04:18.51$\pm$0.11 & 19.0$\pm$3.7 & 15.3$\pm$1.9 & -- & None\\
NGC-2068-B-VLA-75 & 05:46:41.176$\pm$0.01 & -00:04:25.54$\pm$0.02 & 170.4$\pm$5.4 & 151.9$\pm$2.8 & -- & None\\
NGC-2068-B-VLA-76 & 05:46:33.101$\pm$0.07 & -00:04:31.17$\pm$0.11 & 16.6$\pm$3.4 & 15.7$\pm$1.9 & 78-D & None\\
NGC-2068-B-VLA-77 & 05:46:30.831$\pm$0.05 & -00:04:37.22$\pm$0.12 & 26.5$\pm$3.9 & 20.3$\pm$1.8 & 81-D & None\\
NGC-2068-B-VLA-78 & 05:46:33.116$\pm$0.13 & -00:04:40.40$\pm$0.20 & 11.6$\pm$3.8 & 9.5$\pm$1.9 & 76-D & None\\
NGC-2068-B-VLA-79 & 05:46:42.033$\pm$0.14 & -00:04:40.99$\pm$0.28 & 36.7$\pm$9.0 & 16.9$\pm$3.0 & -- & None\\
NGC-2068-B-VLA-80 & 05:46:26.254$\pm$0.09 & -00:04:44.75$\pm$0.13 & 24.2$\pm$4.5 & 16.5$\pm$2.0 & -- & [STS2013] 92011\\
NGC-2068-B-VLA-81 & 05:46:31.065$\pm$0.06 & -00:04:47.48$\pm$0.10 & 31.3$\pm$4.1 & 22.1$\pm$1.8 & 77-D & None\\
NGC-2068-B-VLA-82 & 05:46:34.997$\pm$0.25 & -00:04:55.75$\pm$0.20 & 20.5$\pm$5.7 & 9.5$\pm$1.9 & -- & None\\
NGC-2068-B-VLA-83 & 05:46:31.600$\pm$0.04 & -00:05:00.39$\pm$0.07 & 28.1$\pm$3.3 & 25.2$\pm$1.8 & -- & None\\
NGC-2068-B-VLA-84 & 05:46:33.918$\pm$0.13 & -00:05:15.94$\pm$0.18 & 19.2$\pm$5.2 & 13.8$\pm$2.4 & -- & None\\
NGC-2068-B-VLA-85 & 05:46:38.832$\pm$0.12 & -00:05:20.30$\pm$0.12 & 46.3$\pm$7.6 & 24.2$\pm$2.8 & 86-D & None\\
NGC-2068-B-VLA-86 & 05:46:39.593$\pm$0.13 & -00:05:21.13$\pm$0.14 & 59.4$\pm$9.4 & 24.1$\pm$2.8 & 85-D & None\\
NGC-2068-B-VLA-87 & 05:46:14.214$\pm$0.10 & -00:05:26.64$\pm$0.15 & 68.0$\pm$18.1 & 60.5$\pm$9.6 & -- & HOPS-320; 2MASS J05461423-0005261; HH 22 MIR\\
NGC-2068-B-VLA-88 & 05:46:18.891$\pm$0.02 & -00:05:38.06$\pm$0.03 & 205.7$\pm$11.4 & 187.4$\pm$6.1 & -- & V* V2764 Ori\\
NGC-2068-B-VLA-89 & 05:46:33.332$\pm$0.05 & -00:05:43.88$\pm$0.11 & 32.7$\pm$5.4 & 29.1$\pm$2.8 & -- & None\\
NGC-2068-B-VLA-90 & 05:46:13.147$\pm$0.13 & -00:06:04.69$\pm$0.18 & 71.8$\pm$23.9 & 63.8$\pm$12.7 & -- & HOPS-388; V* V1647 Ori\\
NGC-2068-B-VLA-91 & 05:46:33.017$\pm$0.28 & -00:06:05.81$\pm$0.22 & 30.8$\pm$9.2 & 13.1$\pm$2.9 & -- & None\\
NGC-2068-B-VLA-92 & 05:46:25.477$\pm$0.11 & -00:06:08.07$\pm$0.16 & 21.3$\pm$6.2 & 19.4$\pm$3.3 & -- & None\\
NGC-2068-B-VLA-93 & 05:46:36.809$\pm$0.04 & -00:06:14.71$\pm$0.11 & 132.6$\pm$10.2 & 61.1$\pm$3.3 & -- & None\\
NGC-2068-B-VLA-94 & 05:46:32.775$\pm$0.04 & -00:06:26.64$\pm$0.06 & 58.2$\pm$6.2 & 51.2$\pm$3.3 & -- & None\\
NGC-2068-B-VLA-95 & 05:46:36.872$\pm$0.02 & -00:06:34.86$\pm$0.05 & 169.2$\pm$10.4 & 120.4$\pm$4.7 & -- & None\\
NGC-2068-B-VLA-96 & 05:46:25.634$\pm$0.11 & -00:06:51.43$\pm$0.17 & 40.6$\pm$10.7 & 31.9$\pm$5.2 & -- & None\\
\enddata
\tablecomments{Primary source catalog derived from B-configuration observations taken in 2023.}
\tablenotetext{a}{The source number and VLA configuration where it is blended with another source is listed. For example 2-D, means that the source in that row will be blended with source 2 when observed in D-configuration.}
\end{deluxetable}

\clearpage
\begin{deluxetable}{llllllllll}
\label{per-epoch-fluxes}
\tabletypesize{\tiny}
\tablewidth{0pt}
\tablecaption{Per-Epoch Flux Densities}

\tablehead{\colhead{Date} & \colhead{Julian Date} & \colhead{VLA-14} & \colhead{VLA-14} & \colhead{VLA-15} & \colhead{VLA-15} & \colhead{VLA-48} & \colhead{VLA-48} & \colhead{VLA-88} & \colhead{VLA-88} \\
\colhead{} & \colhead{} & \colhead{Flux Density} & \colhead{N comps} & \colhead{Flux Density} & \colhead{N comps} & \colhead{Flux Density} & \colhead{VLA-48 N comps} & \colhead{Flux Density} & \colhead{N comps} \\
\colhead{} & \colhead{(MJD)} & \colhead{($\mu$Jy)} & \colhead{} & \colhead{($\mu$Jy)} & \colhead{} &\colhead{($\mu$Jy)} & \colhead{} & \colhead{($\mu$Jy)} & \colhead{} \\
}
\startdata
2015-10-06 & 57301.0 & 75.7$\pm$16.2 & 1 & 139.9$\pm$33.8 & 1 & 488.6$\pm$106.2 & 1 & $<$74.0 & 0 \\
2021-03-24 & 59297.0 & 99.4$\pm$14.4 & 1 & 162.8$\pm$41.1 & 1 & 341.9$\pm$55.3 & 1 & $<$54.4 & 0 \\
2021-05-15 & 59349.0 & 81.6$\pm$19.6 & 1 & 185.9$\pm$39.7 & 1 & 353.1$\pm$45.7 & 1 & $<$50.8 & 0 \\
2021-04-20 & 59324.0 & 87.2$\pm$15.3 & 1 & 171.9$\pm$30.0 & 1 & 408.1$\pm$38.2 & 1 & $<$41.7 & 0 \\
2021-07-25 & 59420.0 & 80.9$\pm$21.6 & 1 & 202.5$\pm$61.6 & 1 & $<$78.2 & 0 & $<$55.9 & 0 \\
2021-08-08 & 59434.0 & 76.6$\pm$17.3 & 1 & 112.8$\pm$32.8 & 1 & $<$76.0 & 0 & $<$54.3 & 0 \\
2021-08-24 & 59450.0 & 78.2$\pm$20.3 & 1 & 194.1$\pm$34.0 & 1 & 270.6$\pm$57.7 & 1 & $<$49.0 & 0 \\
2021-09-11 & 59468.0 & 91.0$\pm$13.9 & 1 & 185.0$\pm$49.6 & 1 & 818.5$\pm$55.2 & 1 & $<$47.3 & 0 \\
2021-08-24 & 59450.0 & 77.2$\pm$7.4 & 1 & 173.3$\pm$22.1 & 1 & 330.8$\pm$31.2 & 1 & $<$27.1 & 0 \\
2021-10-15 & 59502.0 & 91.0$\pm$13.7 & 1 & 202.9$\pm$33.8 & 1 & 129.3$\pm$40.0 & 1 & $<$35.0 & 0 \\
2021-11-06 & 59524.0 & 80.1$\pm$13.0 & 1 & 128.3$\pm$26.2 & 1 & 170.3$\pm$39.6 & 1 & $<$35.6 & 0 \\
2021-12-07 & 59555.0 & 79.5$\pm$11.2 & 1 & 99.9$\pm$23.7 & 1 & 812.0$\pm$29.0 & 1 & $<$36.2 & 0 \\
2021-12-30 & 59578.0 & 88.3$\pm$13.6 & 1 & 135.4$\pm$20.9 & 1 & 531.6$\pm$32.6 & 1 & 67.4$\pm$19.9 & 1 \\
2022-01-12 & 59591.0 & 103.5$\pm$9.2 & 1 & 124.4$\pm$19.6 & 1 & 1312.2$\pm$26.5 & 1 & 79.5$\pm$20.8 & 1 \\
2021-11-30 & 59548.0 & 85.6$\pm$5.0 & 1 & 134.0$\pm$12.2 & 1 & 660.7$\pm$18.3 & 1 & 32.6$\pm$9.9 & 1 \\
2022-03-05 & 59643.0 & 69.5$\pm$16.0 & 1 & 186.1$\pm$45.7 & 1 & 137.6$\pm$30.5 & 1 & $<$34.8 & 0 \\
2022-04-07 & 59676.0 & 68.9$\pm$13.1 & 1 & 126.7$\pm$33.6 & 1 & 1516.5$\pm$58.2 & 2 & $<$33.8 & 0 \\
2022-05-21 & 59720.0 & 109.6$\pm$13.8 & 1 & 145.1$\pm$27.1 & 1 & 165.0$\pm$36.5 & 1 & $<$31.7 & 0 \\
2022-06-24 & 59754.0 & 86.2$\pm$15.2 & 1 & 176.9$\pm$36.6 & 1 & 276.5$\pm$30.7 & 1 & $<$33.1 & 0 \\
2022-05-01 & 59700.0 & 80.8$\pm$7.1 & 1 & 150.5$\pm$18.5 & 1 & 422.8$\pm$23.4 & 1 & $<$16.8 & 0 \\
2022-07-23 & 59783.0 & 94.9$\pm$18.1 & 1 & 220.7$\pm$45.0 & 1 & 249.4$\pm$41.7 & 1 & $<$53.4 & 0 \\
2022-08-20 & 59811.0 & 101.6$\pm$22.3 & 1 & 168.7$\pm$49.8 & 1 & 378.4$\pm$48.5 & 1 & $<$62.7 & 0 \\
2022-09-20 & 59842.0 & 99.4$\pm$20.6 & 1 & 233.9$\pm$42.8 & 1 & 1907.2$\pm$41.6 & 1 & $<$58.0 & 0 \\
2022-08-20 & 59811.0 & 94.5$\pm$13.1 & 1 & 178.1$\pm$25.7 & 1 & 934.2$\pm$32.2 & 1 & $<$42.7 & 0 \\
2022-09-30 & 59852.0 & 104.3$\pm$13.7 & 1 & 195.2$\pm$34.1 & 1 & 849.0$\pm$44.5 & 1 & $<$41.6 & 0 \\
2022-11-03 & 59886.0 & 127.4$\pm$16.5 & 1 & 154.3$\pm$31.6 & 1 & 614.2$\pm$53.7 & 1 & $<$41.8 & 0 \\
2022-12-01 & 59914.0 & 149.8$\pm$14.8 & 1 & 197.1$\pm$41.0 & 1 & 1498.6$\pm$43.9 & 1 & 198.5$\pm$30.8 & 1 \\
2022-12-27 & 59940.0 & 147.4$\pm$15.6 & 1 & 271.7$\pm$51.0 & 1 & 259.4$\pm$48.8 & 1 & $<$39.2 & 0 \\
2022-11-15 & 59898.0 & 134.4$\pm$9.1 & 1 & 198.4$\pm$24.6 & 1 & 737.3$\pm$26.5 & 1 & 67.1$\pm$11.4 & 1 \\
2023-01-17 & 59961.0 & 134.5$\pm$9.4 & 1 & 131.7$\pm$24.3 & 1 & 575.5$\pm$26.5 & 1 & $<$28.1 & 0 \\
2023-03-21 & 60024.0 & 126.2$\pm$11.8 & 1 & 135.9$\pm$29.7 & 1 & 309.5$\pm$36.4 & 1 & $<$38.9 & 0 \\
2023-03-29 & 60032.0 & 128.3$\pm$11.0 & 1 & 203.9$\pm$28.1 & 1 & 755.9$\pm$32.2 & 1 & $<$36.6 & 0 \\
2023-04-02 & 60036.0 & 142.6$\pm$9.1 & 1 & 107.2$\pm$20.3 & 1 & 258.4$\pm$28.0 & 1 & $<$26.9 & 0 \\
2023-04-30 & 60064.0 & 107.9$\pm$11.0 & 1 & 125.6$\pm$24.6 & 1 & 210.1$\pm$36.7 & 1 & 852.8$\pm$22.8 & 1 \\
2023-05-04 & 60068.0 & 115.0$\pm$14.1 & 1 & 153.7$\pm$34.2 & 1 & 717.2$\pm$34.6 & 1 & 848.4$\pm$27.0 & 1 \\
2023-05-16 & 60080.0 & 118.2$\pm$11.1 & 1 & 178.8$\pm$37.7 & 1 & 235.9$\pm$54.9 & 1 & 149.4$\pm$27.0 & 1 \\
2023-05-19 & 60083.0 & 106.8$\pm$13.0 & 1 & 135.3$\pm$35.7 & 1 & 270.5$\pm$38.6 & 1 & 170.4$\pm$28.4 & 1 \\
2023-03-30 & 60033.0 & 125.6$\pm$4.1 & 1 & 138.1$\pm$11.3 & 1 & 406.6$\pm$12.9 & 1 & 205.7$\pm$11.4 & 1 \\
2023-08-07 & 60163.0 & 138.1$\pm$11.3 & 1 & 162.5$\pm$44.4 & 1 & 286.8$\pm$35.3 & 1 & $<$32.9 & 0 \\
2023-08-11 & 60167.0 & 154.6$\pm$7.0 & 1 & 128.3$\pm$25.4 & 1 & 215.8$\pm$22.5 & 1 & $<$22.5 & 0 \\
2023-08-19 & 60175.0 & 174.4$\pm$13.1 & 1 & 221.5$\pm$44.3 & 2 & 167.3$\pm$28.8 & 1 & $<$31.8 & 0 \\
2023-08-24 & 60180.0 & 169.8$\pm$13.9 & 1 & 115.1$\pm$33.1 & 1 & 161.0$\pm$39.9 & 1 & $<$35.4 & 0 \\
2023-08-27 & 60183.0 & 162.4$\pm$10.1 & 1 & 109.0$\pm$28.8 & 1 & 169.7$\pm$32.1 & 1 & $<$32.1 & 0 \\
2023-09-17 & 60204.0 & 153.2$\pm$11.6 & 1 & 188.9$\pm$44.4 & 1 & 580.4$\pm$46.1 & 2 & $<$29.3 & 0 \\
2023-08-24 & 60180.0 & 154.4$\pm$3.9 & 1 & 151.3$\pm$18.2 & 1 & 237.4$\pm$11.6 & 1 & $<$12.3 & 0 \\
2026-02-22 & 61093.0 & 198.0$\pm$14.9 & 2 & 109.8$\pm$20.3 & 1 & 156.1$\pm$21.9 & 1 & $<$24.0 & 0 \\

\enddata
\tablecomments{Only a subset of all flux densities are reported here, the rest are provided as a machine readable table.}
\end{deluxetable}

\end{document}